\documentclass[prb,showpacs,aps,twocolumn]{revtex4-1}
\usepackage{graphicx}% Include figure files
\usepackage{epstopdf}
\usepackage{bm}% bold math
\usepackage{color}
\usepackage{ulem}
\usepackage{amsfonts,amsthm}
\usepackage{amsmath}
\usepackage{amssymb}

\begin{document}
\newcommand{\tr}[1]{\textcolor{red}{#1}}

\newcommand{\avrg}[1]{\left\langle #1 \right\rangle}
\newcommand{\eqsa}[1]{\begin{eqnarray} #1 \end{eqnarray}}
\newcommand{\eqwd}[1]{\begin{widetext}\begin{eqnarray} #1 \end{eqnarray}\end{widetext}}
\newcommand{\hatd}[2]{\hat{ #1 }^{\dagger}_{ #2 }}
\newcommand{\hatn}[2]{\hat{ #1 }^{\ }_{ #2 }}
\newcommand{\wdtd}[2]{\widetilde{ #1 }^{\dagger}_{ #2 }}
\newcommand{\wdtn}[2]{\widetilde{ #1 }^{\ }_{ #2 }}
\newcommand{\cond}[1]{\overline{ #1 }_{0}}
\newcommand{\conp}[2]{\overline{ #1 }_{0#2}}
\newcommand{\nn}{\nonumber\\}
\newcommand{\cdt}{$\cdot$}
\newcommand{\bra}[1]{\langle#1|}
\newcommand{\ket}[1]{|#1\rangle}
\newcommand{\braket}[2]{\langle #1 | #2 \rangle}
\newcommand{\bvec}[1]{\mbox{\boldmath$#1$}}
\newcommand{\blue}[1]{{#1}}
\newcommand{\bl}[1]{{#1}}
\newcommand{\bn}[1]{\textcolor{blue}{#1}}
\newcommand{\rr}[1]{{#1}}
\newcommand{\bu}[1]{\textcolor{blue}{#1}}
\newcommand{\red}[1]{{#1}}
\newcommand{\fj}[1]{{#1}}
\newcommand{\green}[1]{{#1}}
\newcommand{\gr}[1]{\textcolor{green}{#1}}
\definecolor{green}{rgb}{0,0.5,0.1}
\definecolor{blue}{rgb}{0,0,0.8}
\preprint{APS/123-QED}

\title{
Phase Diagram Structure of Topological Mott Transition for Zero-gap Semiconductors beyond Conventional Landau-Ginzburg-Wilson Scenario
}
%\if0
\author{Moyuru Kurita}
\author{Youhei Yamaji}
\author{Masatoshi Imada}
\affiliation{Department of Applied Physics, University of Tokyo, Hongo, Bunkyo-ku, Tokyo, 113-8656, Japan.}%
\affiliation{CREST, JST, Hongo, Bunkyo-ku, Tokyo, 113-8656, Japan.
}%
\date{December 29, 2011}% It is always \today, today,

\begin{abstract}
We show that a wide class of unconventional quantum criticality emerges
when orbital currents cause quantum phase transitions from
zero-gap semiconductors such as Dirac fermions to
topological insulator (TI) or Chern insulator (CI).
Changes in Fermi surface topology concomitant with (SU(2) or time reversal) symmetry breakings generate {\it quantum critical lines} (QCL) even beyond the quantum critical point.
This QCL running at temperature $T=0$ separates two distinct topological phases.
This is in contrast to the simple termination of the finite temperature critical line at the quantum critical point without any extension of it at $T=0$.
Topology change causes the unconventionality beyond the concept of simple spontaneous symmetry breaking assumed in the conventional Landau-Ginzburg-Wilson (LGW) scenario.
The unconventional universality implied by 
mean-field critical exponents $\beta>1/2$ and $\delta<3$ is protected by the existence of the quantum critical line.
It emerges for several specific lattice models including the honeycomb, kagome, diamond
and pyrochlore lattices.
We also clarify phase diagrams of the topological phases in these lattices at finite temperatures.

\end{abstract}

%05.30.Rt Quantum phase transition
%71.10.Fd Hubbard model electronic structure
%73.43.Lp quantum Hall effects
%71.27.+a Strongly correlated electron systems

\maketitle

\section{Introduction}

Critical phenomena of
phase transitions
are classified into a small number of universality classes.
For such symmetry-breaking transitions,
the LGW scheme
is extremely successful~\cite{Landau,Wilson}.
Usual Fermi sea
of electrons in crystalline solids, a {\it vacuum} for particle-hole excitations,
is 
known to be unstable
in the presence of electron-electron interactions, leading to
spontaneous
symmetry breakings. 
Competitions of different %types of such 
spontaneous symmetry breakings,
for instance, magnetism and superconductivity are the subject of 
extensive studies on strongly correlated electron systems.
Although it has a rich diversity, most phase transitions 
accompanying the spontaneous symmetry breakings are primarily described by the 
LGW framework~\cite{Landau,Wilson} for quantum phase transitions.  
In spite of its great success,
however, the LGW scheme has recently been challenged
from various viewpoints, especially in strongly correlated electrons
\cite{Belitz,Senthil,Imada}. It should also be noted that changes in topology such as those of Fermi-surface topology as the Lifshitz transition constitute a class of quantum phase transitions distinct from the LGW scheme \cite{Imada,XGWen,Yamaji}. 

Contrary to the conventional Fermi sea,
particle-hole excitations %in solids 
sometimes behave
as massless Dirac fermions
%~\cite{Haldane} 
as
realized in graphene~\cite{CastroNeto}
and at surfaces of TI~\cite{Kane_Mele,Moore,Roy,Hasan}.
These {\it zero-gap semiconductors}
attract much attention because of peculiar transport
properties, such as
ballistic transport~\cite{Ando} and quantum Hall effects~\cite{Haldane,Novoselov}.
Unlike the usual Fermi sea, %instabilities,
band crossing points (Fermi points of the zero-gap semiconductors) 
are 
protected by certain symmetries~\cite{Kane_Mele,Ryu} 
and thus are inferred to have weaker instabilities
%[Really written in the Ref.12?]}~
\cite{Ando,Gonzalez}.

However, 
the zero-gap semiconductors are sensitive to the spin-orbit interaction,
and time-reversal-symmetry broken fields such as magnetic flux.
These stabilize topologically nontrivial phases as TI 
for the former and quantum Hall insulators (or CI)~\cite{Haldane} for the latter. 
It induces microscopic loop of spin (or charge) currents, for which we inclusively call {\it orbital currents}. 
We call phase transitions between the orbital-current phases (TI or CI) and 
zero-gap semiconductors orbital-current transition (OCT).

Even without such external triggers,
electron-electron interactions are proposed to cause spontaneous symmetry breakings
generating the orbital currents 
\cite{Raghu,Zhang_Ran,Wen,Kurita}.
This spontaneous-orbital-current phase is called a topological Mott insulators (TMI).
We call transitions to TMI, topological Mott transitions(TMT).  

In this Paper we propose that
OCT including TMT shows unprecedented properties, because of the involvement of the topological change in the Fermi surface inevitably induced simultaneously with the spontaneous symmetry breaking, beyond the concept of the simple symmetry breaking, which causes non-LGW type quantum phase transitions characterized by
(1) {\it Unconventional critical exponents} $\beta > 1/2$ and $\delta < 3$, and
(2) the existence of {\it QCL} representing semimetal-topological insulator transition at $T = 0$ extending for $V < V_c$, where $T$ is temperature and $V$ is a controlling parameter. Here, we note that the conventional quantum criticality induced by the enhanced quantum fluctuation leading to vanishing critical temperature does not have such a QCL extending beyond the quantum critical point if the phase transition is caused simply by the spontaneous symmetry breaking.
We generally show that unconventional universality arises depending on spatial dimension and band-dispersion exponent near the Fermi level at $T=0$.
We also focus on several choices of lattice models to apply our general theory and study the criticality at finite temperatures.
This opens new avenues of studies on quantum critical phenomena 
and provides a clue for potential applications of TI (or CI) and its transitions.

\section{GENERAL THEORY}\label{sec:general theory}

\subsection{Models}\label{sec:models}

Zero-gap semiconductors are in general described by a simple dispersion 
$\epsilon_{+}$ and $\epsilon_{-}$
as functions of momenta $k$,
\begin{eqnarray}
	\epsilon_{\pm} = \pm v_{\pm} k^{n},
\end{eqnarray}
with a double degeneracy at $k=0$.
Here, $n=1$ for Dirac fermions and $n=2$ for quadratic band crossings.

\begin{figure}
\centering
\includegraphics[width=8.5cm]{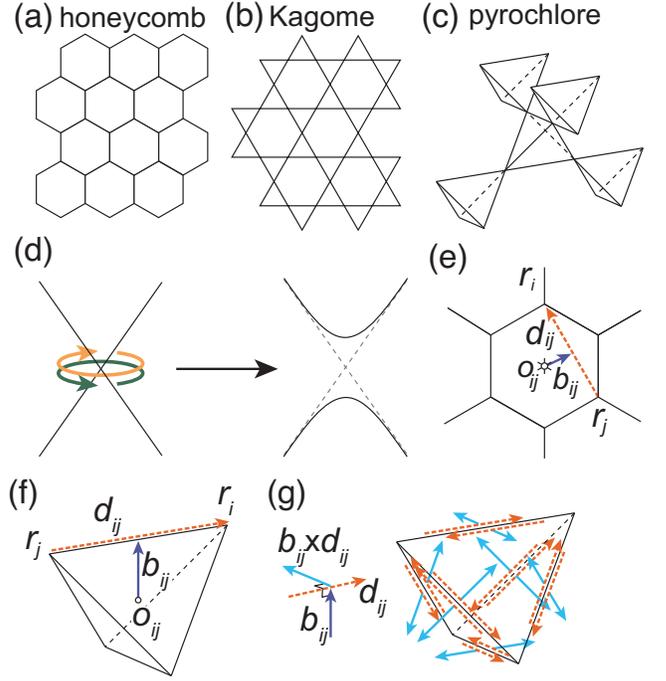}
\caption{
(Color online) (a) Honeycomb, (b) kagom${\rm \acute{e}}$ and (c) pyrochlore lattice structures.
(d) Schematics showing how degeneracy at a band-crossing point is lifted.
The loop current in this momentum space is equivalent to that in the real space shown in (g). 
Spin-orbital-current configurations on a hexagon (e) and a tetrahedron (f): Along each nearest neighbor bond $(r_i,r_j)$, opposite spins on the directions $\pm \bm{b}_{ij}\times \bm{d}_{ij}$ flow in the opposite directions as illustrated in (g).
Here, we define $\bm{d}_{ij}=\bm{r}_{i}-\bm{r}_{j}$ and $\bm{b}_{ij}=(\bm{r}_{i}+\bm{r}_{j})/2-\bm{o}_{ij}$ with $\bm{o}_{ij}$ being the gravity center of the hexagon (e) (tetrahedron (f)).
In (g), all the vectors schematically illustrate only the correct directions while the lengths are not correct.
\label{fig:K_fig_1}}
\end{figure}

While Dirac fermions appear, for example, in honeycomb \cite{Haldane} and diamond lattices \cite{Zhang_Ran}, quadratic band-crossing points show up in kagom${\rm \acute{e}}$~\cite{Sun,Wen}, and pyrochlore lattices \cite{Guo,Kurita} (Fig.\ref{fig:K_fig_1}(a)-(c)), for the single-band systems, where, only in the pyrochlore lattice, three bands touch each other at the band-crossing point, instead of two.

The orbital currents make the band-crossing point (Fig.\ref{fig:K_fig_1}(d)) in the simplest case as,
\begin{eqnarray}
	\epsilon_{\pm} =\pm v_{\pm} k^{n} \rightarrow \pm \sqrt{m^2 + v_{\pm}^2 k^{2n}}.
\end{eqnarray}
A mass $m$ is scaled by, as we will discuss later, the orbital current $\zeta$ multiplied by intersite Coulomb repulsion $V$; $m\propto V\zeta$.
Orbital currents are sketched in Fig.\ref{fig:K_fig_1}(e)-(g).

The origin of the gap formation is understood from the lifting of the degeneracy related to the lattice symmetry.
For example, in 2D lattices like the honeycomb and kagom${\rm \acute{e}}$ lattices, the clock- and counterclock-wise rotating electrons are doubly degenerating at the band-crossing point.
The orbital currents break the symmetry, and lift the 2-fold degeneracy between them, leading to a gap (Fig.\ref{fig:K_fig_1}(d)).

To understand the criticality of OCT, especially at $T=0$, we expand the free energy $F$ in terms of the order parameter $\zeta$ based on a microscopic hamiltonian and realistic band dispersions.
Although we formally follow the spirit of LGW expansion, we will see later that $F$ has a singularity expressed by $|\zeta|^{d/n +1}$ which is different from the conventional LGW result.

As a microscopic Hamiltonian, we study an extended Hubbard model,
\begin{eqnarray}
	H &=& H_{0} + U\sum_{i}n_{i\uparrow}n_{i\downarrow} + \sum_{i,j}\frac{V_{ij}}{2}n_in_j \notag \\
	H_{0} &=& -t\sum_{\langle i,j \rangle \sigma} c^{\dagger}_{i\sigma}c_{j\sigma},
	\label{Eq.2}
\end{eqnarray}
where $c^{\dagger}_{i\sigma}$ ($c_{i\sigma}$) is a creation (annihilation) operator for a $\sigma$-spin electron, $n_i=n_{i\uparrow}+n_{i\downarrow}$ is an electron density operator, $\langle i,j \rangle$ is a pair of nearest-neighbor sites, and $U$ ($V_{ij}$) are an on-site (off-site) Coulomb repulsion.
We only consider the nearest ($V_{ij}=V_1$) and 2nd ($V_{ij}=V_2$) neighbor interactions for $V_{ij}$, for simplicity.

\subsection{Orbital-current mean field}\label{sec:orbital-current}

\begin{figure}
\centering
\includegraphics[width=8.5cm]{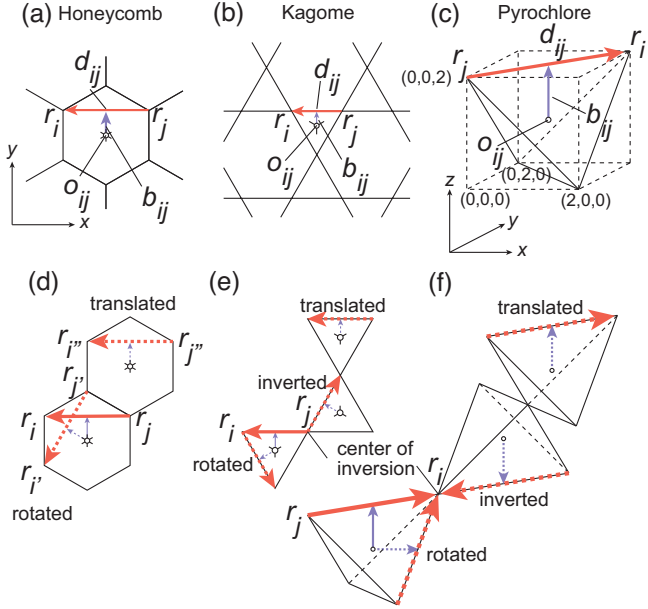}
\caption{
(Color online)
(a)-(c)
Examples for vectors $\bm{r}_{i}$, $\bm{r}_{j}$, $\bm{o}_{ij}$,
$\bm{b}_{ij}$, and $\bm{d}_{ij}$ for the honeycomb, kagom${\rm \acute{e}}$ and pyrochlore
lattice.
(d)-(f) Examples of symmetric operations applied to a bond $(i,j)$.
\label{fig:Fig01}}
\end{figure}

The Hartree-Fock decoupling of the interaction ($V$) term in Eq.(\ref{Eq.2}) is naturally described by a product of two spontaneous orbital currents that preserve lattice symmetries\cite{Raghu}. 
Among possible mean fields, two types of loop orbital currents exist: One is a mean field of the spin-orbital current,
\begin{eqnarray}
	\langle c^{\dagger}_{j\beta}c_{i\alpha} \rangle &=& g\sigma_{0\alpha\beta} -i\zeta_s \frac{\bm{b}_{ij}\times \bm{d}_{ij}}{|\bm{b}_{ij}\times \bm{d}_{ij}|}\cdot \bm{{\sigma}}_{\alpha\beta}, \notag \\
	\zeta_s &=& \frac{i}{2} \sum_{\alpha,\beta} \langle c^{\dagger}_{j\beta}c_{i\alpha} \rangle \frac{\bm{b}_{ij}\times \bm{d}_{ij}}{|\bm{b}_{ij}\times \bm{d}_{ij}|}\cdot \bm{\sigma}_{\beta\alpha},
	\label{Eq.3}
\end{eqnarray}
which creates time-reversal-symmetric topological insulators and the other is a mean field of the charge-orbital current,
\begin{eqnarray}
	\langle c^{\dagger}_{j\sigma}c_{i\sigma} \rangle &=& g -i\zeta_c
	\left[ \frac{\bm{b}_{ij}\times \bm{d}_{ij}}{|\bm{b}_{ij}\times \bm{d}_{ij}|} \right]_{z}, \notag \\
	\zeta_c &=& \frac{i}{2} \sum_{\sigma}
	\langle c^{\dagger}_{j\sigma}{c}_{i\sigma} \rangle
	\left[ \frac{\bm{b}_{ij}\times \bm{d}_{ij}}{|\bm{b}_{ij}\times \bm{d}_{ij}|} \right]_{z},
	\label{Eq.4}
\end{eqnarray}
which induces time-reversal-symmetry broken Chern insulators. 
For definitions of $\bm{b}_{ij}$ and $\bm{d}_{ij}$, see Fig.\ref{fig:K_fig_1}(g) and Fig.\ref{fig:Fig01} (a)-(c).
We note that the form (\ref{Eq.3}) is identical to the spin-orbit interaction uniquely allowed by the lattice symmetry and the time-reversal invariance.

Here, we show that the spin-orbit interaction as well as the orbital-current mean field has a unique form if we preserve the lattice and time-reversal symmetries.
Let us first consider some symmetry operations satisfied by honeycomb, kagom${\rm \acute{e}}$, and pyrochlore lattices.
When we pick up a bond between $i$-th ($\bm{r}_{i}$) and $j$-th sites ($\bm{r}_{j}$) of a lattice,  the system is invariant under 3 important symmetric operations related to the bond: 
\begin{enumerate}
\item The time reversal operation $\Theta$.
\item A reflection with respect to a plane perpendicular to the bond and containing the center of the bond, which is denoted by $R_{(i,j)}^{\perp}$.
\item A $\pi$-rotation around the axis perpendicular to the bond and crossing at the center of the bond each other, which is denoted as $R_{(i,j)}^{\pi}$.
\end{enumerate}
For all the lattices considered, the axis for $R_{(i,j)}^{\pi}$ that keeps the lattice invariant is along the vector $\bm{b}_{ij}$ pointing to the center of the bond $(i,j)$ from the gravity center $\bm{o}_{ij}$ of the unit cell containing $i$ and $j$ (see Fig.\ref{fig:Fig01}).
This is defined as $\bm{b}_{ij}=(\bm{r}_{i}+\bm{r}_{j})/2-\bm{o}_{ij}$.
For the honeycomb, kagom${\rm \acute{e}}$ and pyrochlore lattices, $\bm{o}_{ij}$ is the gravity center of the hexagon, triangle and tetrahedron, respectively, as illustrated in Fig.\ref{fig:Fig01}(a)-(c).

Next we introduce a current flowing through the bond $(i,j)$ along $\bm{d}_{ij}=\bm{r}_{i}-\bm{r}_{j}$, described by the hermitian charge/spin current operator $J^{(a)}_{(i,j)}(=[J^{(a)}_{(i,j)}]^{\dagger})$ defined by
\begin{eqnarray}
	J^{(a)}_{(i,j)} =(-i) \sum_{\alpha,\beta = \uparrow, \downarrow}
	\left[ c^{\dagger}_{i\alpha} \sigma_{a\alpha\beta} c_{j\beta} - c^{\dagger}_{j\alpha} \sigma_{a\alpha\beta} c_{i\beta} \right].
	\label{CO_0xyz}
\end{eqnarray}
Here, a $2 \times 2$ matrix $\sigma_{a}$ represents, for $a=0$, a 2D identity matrix describing the charge current or, for $a=x,y,z$, Pauli matrices describing the spin currents.

Let us examine symmetric properties of $J^{(a)}_{(i,j)}$.
As charge currents break the time-reversal symmetry while spin currents preserve it, $J^{(0)}_{(i,j)}$ and $J^{(x,y,z)}_{(i,j)}$ are time-reversal odd and even, respectively.

Similarly, the reflection $R_{(i,j)}^{\perp}$ and the rotation $R_{(i,j)}^{\pi}$ transform $J^{(0)}_{(i,j)}$ into
\begin{eqnarray}
	R_{(i,j)}^{\perp}[J^{(0)}_{(i,j)}] =R_{(i,j)}^{\pi}[J^{(0)}_{(i,j)}] = -J^{(0)}_{(i,j)}.
\end{eqnarray}
Therefore we note that, if the lattice symmetry satisfies these reflection and rotation as symmetric operations, the charge-current mean field breaks the reflection/rotation symmetry in addition to the time-reversal symmetry.
When the charge currents form a loop, after setting currents on every bond as we will discuss below, these currents break the so-called chiral symmetry.

On the other hand, spin current operators $J^{(x,y,z)}_{(i,j)}$ are transformed in complicated ways: The reflection $R_{(i,j)}^{\perp}$ makes a spin rotate $\pi$ around $\bm{d}_{ij}$, while $R_{(i,j)}^{\pi}$  makes the spin rotate $\pi$ around $\bm{b}_{ij}$.
In addition, the both operations exchange the sites $i$ and $j$.
For example, if we take
\begin{eqnarray}
	\bm{d}_{ij}=(-|\bm{d}_{ij}|,0,0) \quad \mathrm{and} \quad \bm{b}_{ij}=(0,|\bm{b}_{ij}|,0),
\end{eqnarray}
the reflection $R_{(i,j)}^{\perp}$ transforms a vector $(J^{(x)}_{(i,j)},J^{(y)}_{(i,j)},J^{(z)}_{(i,j)})$ into $(-J^{(x)}_{(i,j)},J^{(y)}_{(i,j)},J^{(z)}_{(i,j)})$, while the rotation $R_{(i,j)}^{\pi}$ transforms it into $(J^{(x)}_{(i,j)},-J^{(y)}_{(i,j)},J^{(z)}_{(i,j)})$.
Thus, if a spin current preserves the two symmetries, the spin alignment vector $\bm{n}_{(i,j)}$ of the current from $j$-th to $i$-th sites, has to be along the $z$-axis.
More generally, when the above three symmetries are preserved, the unit vector $\bm{n}_{(i,j)}$ has to be perpendicular to both of $\bm{d}_{ij}$ and $\bm{b}_{ij}$ and uniquely has the form
\begin{eqnarray}
	\bm{n}_{(i,j)}= \pm \frac{\bm{b}_{ij} \times \bm{d}_{ij}}{|\bm{b}_{ij}\times \bm{d}_{ij}|}.
\end{eqnarray}

Now, the spin current operator is expressed by 
\begin{eqnarray}
	&&J^{(s)}_{(i,j)}(\bm{n}_{(i,j)}) = n^{(x)}_{(i,j)}J^{(x)}_{(i,j)}+n^{(y)}_{(i,j)}J^{(y)}_{(i,j)}+n^{(z)}_{(i,j)}J^{(z)}_{(i,j)}
	\nonumber\\
	&=& (-i)\sum_{\alpha, \beta} \left[ c^{\dagger}_{i\alpha} \{ \bm{n}_{(i,j)}\cdot \bm{\sigma} \}_{\alpha\beta} c_{j\beta}- c^{\dagger}_{j\alpha} \{ \bm{n}_{(i,j)}\cdot \bm{\sigma} \}_{\alpha\beta} c_{i\beta} \right].
	\nonumber\\
\end{eqnarray}

We also note that for the honeycomb lattice, the bond $(i,j)$ has to be between the 2nd neighbor sites, because the nearest-neighbor and the 3rd-neighbor bonds have an additional $\pi$-rotation symmetry with the axis crossing the honeycomb plane perpendicularly at the center of the bond.
This symmetry prohibits the currents even in the $\bm{b}_{ij}\times\bm{d}_{ij}$ direction. 
In a similar way, for the kagom${\rm \acute{e}}$ and pyrochlore lattices, $\bm{d}_{ij}\neq \bm{0}$ $(i,j)$ has to be between the nearest-neighbor sites. (See Fig.\ref{fig:Fig01}(a)-(c).)

After choosing the direction and spin alignment of the flow, we apply a symmetry operation, which belongs to the point group of the underlying lattice, to the bond $(i,j)$.
For example, in the honeycomb lattice, every 2nd-neighbor bond, where orbital currents flow, is involved in a hexagon. A 6-fold rotation ($C_{6}$ rotation), around the center of such a hexagon, $\bm{o}_{ij}$, is one of the examples of symmetric operations.
If we apply the $C_{6}$ rotation, we obtain the bond $(i',j')$ shown in Fig.\ref{fig:Fig01}(d).
After successive applications of C$_6$ rotations around $\bm{o}_{ij}$, we can determine the direction and spin alignment of the flow of 5 more bonds involved in the hexagon.
Under these operations, $\bm{o}_{ij}$ is invariant, and $\bm{b}_{ij}$ and $\bm{d}_{ij}$ are transformed into, for example, $\bm{b}_{i'j'}$ and $\bm{d}_{i'j'}$, respectively.
If we further use the translational invariance of the honeycomb lattice, we can determine the direction and spin alignment of the flow for all of the 2nd neighbor bonds on the lattice.
For example, we obtain the bond $(i",j")$ by translating the bond $(i,j)$ shown in Fig.\ref{fig:Fig01}(d).
Simultaneously, vectors $\bm{b}_{ij}$ and $\bm{d}_{ij}$ are transformed into $\bm{b}_{i"j"}$ and $\bm{d}_{i"j"}$ as well.
When the unit cells share their corner, for a kagom${\rm \acute{e}}$ and pyrochlore lattice, the inversion operations shown in Fig.\ref{fig:Fig01}(e) and (f) can be used.

Then an explicit expression for mean fields of orbital current with a magnitude $\zeta$, together with a hopping renormalization $g$ due to a usual Fock term is obtained from Eq.(\ref{CO_0xyz}) for the mean field of spin-orbital current 
\begin{eqnarray}
	\langle c^{\dagger}_{j\beta}c_{i\alpha} \rangle = \left( g \sigma_{0} - i\zeta_s \frac{\bm{b}_{ij}\times \bm{d}_{ij}}{|\bm{b}_{ij}\times \bm{d}_{ij}|}\cdot \bm{\sigma} \right)_{\alpha\beta},
	\label{SOC}
\end{eqnarray}
which creates the topological insulators, or a mean field of charge-orbital current,
\begin{eqnarray}
	\langle c^{\dagger}_{j\beta}c_{i\alpha} \rangle =
	\left( g - i\zeta_c \left( \frac{\bm{b}_{ij}\times \bm{d}_{ij}}{|\bm{b}_{ij}\times \bm{d}_{ij}|} \right)_{z} \right) \sigma_{0\alpha\beta},
	\label{COC}
\end{eqnarray}
which induces the Chern insulators (absent in 3D lattices).
These mean fields are generated through a decoupling of the two-body interaction term,
\begin{eqnarray}
	&&n_{i\alpha}n_{i\beta} \rightarrow
	\langle n_{i\alpha} \rangle n_{j\beta} + n_{i\alpha} \langle n_{j\beta} \rangle
	- \langle n_{i\alpha} \rangle \langle n_{j\beta} \rangle
	\nonumber\\
	&-&
	\langle c^{\dagger}_{i\alpha}c_{j\beta} \rangle c^{\dagger}_{j\beta}c_{i\alpha} - c^{\dagger}_{i\alpha}c_{j\beta}
	\langle c^{\dagger}_{j\beta}c_{i\alpha} \rangle + \left| \langle c^{\dagger}_{i\alpha}c_{j\beta} \rangle \right|^{2}.
\end{eqnarray}

Here we note that the orbital-currents instabilities compete with charge and/or spin density wave instabilities in real lattice structures.
Many studies were carried out to search the conditions that an orbital-current phase becomes a stable mean-field ground state.
In the plane of $U,V$ and $t$, phase diagrams clarified on various lattices after considering the competition indeed contain the region of stable symmetry breaking into either $\zeta_s\ne 0$ or $\zeta_c\ne0$ phases within realistic parameters\cite{Raghu,Wen,Kurita}. 

\subsection{Free energy expansion}

The resultant free energy $f[\zeta]$ for this region is essentially expanded by the orbital-current order parameter $\zeta$ (either $\zeta_s$ or $\zeta_c$ in the same way) as 
\begin{eqnarray}
	f[\zeta]-f[0] = -\lambda \zeta + a\zeta^2 +b_{\pm}f_{\rm s}[\zeta] +({\rm higher\ order}),
	\label{Eq.5}
\end{eqnarray}
for small $\zeta$ close to OCT.
Here $a$ and $b_{\pm}$ are constants. 
We add a spin-orbit coupling or magnetic flux $\lambda$ conjugate to $\zeta$, as a straightforward analogue of magnetic fields in magnetic phase transitions.
As we sill show later, in addition to the regular term proportional to $a$, the singular part $f_{\rm s}[\zeta] \propto |\zeta|^{d/n +1}$ emerges at $T=0$ with possible logarithmic corrections. 
The coefficient $b_{+}$ for $\zeta > 0$ is not necessarily equal to $b_-$ for $\zeta < 0$. 
The expansion Eq.(\ref{Eq.5}) is only piecewise analytic separately in $\zeta>0$ and $\zeta<0$ with a nonanalyticity at $\zeta=0$, originating from topological nature of this transition, in contrast to analytic expansions assumed in the LGW framework.

Using the decoupling into orbital-current mean fields given in Eqs.(\ref{SOC}) and (\ref{COC}), the mean-field Hamiltonians for specific lattices are derived.
This orbital-mean-field state satisfies the self-consistency condition and is stabilized for an appropriate choice of parameters, for example, $(U, V_1, V_2)$\cite{Raghu,Wen,Zhang_Ran}. 
In this section, we derive a general and simple analytic formulae of free-energy expansions for doubly-degenerate band crossings, by focusing on low-energy band dispersion around band-crossing points.
It is applied to the case of the honeycomb and kagom${\rm \acute{e}}$ lattices (see also Sec. \ref{sec:Honeycomb lattice} and \ref{sec:Kagome lattice}).
On the other hand, free energy expansion for a pyrochlore lattice which has a triply-degenerate band crossing and a peculiar inversion of valence and conduction bands occurring by the sign change of the spin-orbital current, Ā, requires slightly different considerations.
We will focus on this case in Sec. \ref{sec:Pyrochlore lattice}.
Here, we note that the transition discussed in this article is different from that between the topological and band insulators discussed in Ref.\onlinecite{Murakami1},\onlinecite{Murakami2}, where the phase transitions are driven not by mean fields of the electron correlation but by other external parameters.

When the magnitude of an orbital current $\zeta$ between the $\alpha$-th neighbor sites becomes nonzero, two bands split in the momentum space, $\epsilon_{\pm}(\zeta,\bm{k})$, with a gap scaled by $|\zeta|$.
Then we obtain a mean-field free energy per unit cell as,
\begin{eqnarray}
	f[\zeta] = \frac{F[\zeta]}{N_{u}} &=& - \frac{2T}{N_{u}}\sum_{\bm{k}} \left\{ \ln \left[1+e^{-(\epsilon_{+}(\zeta,\bm{k})-\mu)/T}\right] \right. \notag \\
	&+& \left. \ln \left[1+e^{-(\epsilon_{-}(\zeta,\bm{k})-\mu)/T}\right] \right\} + z_{\alpha}n_{\rm u}V_{\alpha}\zeta^2, \notag \\
	\label{genF}
\end{eqnarray}
where $N_{u}$ is the number of unit cells, $z_{\alpha}$ is the coordination number of the $\alpha$-th neighbor site, $n_{\rm u}$ is the number of sites in a unit cell, and $V_{\alpha}$ is the Coulomb repulsion between the $\alpha$-th neighbor sites.
The chemical potential $\mu$ is chosen to keep the electron density or filling at a fixed value to keep the valence band $\epsilon_{-}$ fully-filled at $T=0$.

Here we focus on a small energy window around a band crossing point $\bm{k}_{\rm BC}$, and introduce a cutoff $\Lambda$, which does not change any essential physics at low temperatures.
If we choose a proper $\Lambda$ for each lattice, we will show later that full numerical results for the original microscopic Hamiltonian are quantitatively reproduced.
Then we expand the band dispersion $\epsilon_{\pm}$ with respect to the momentum measured from the band crossing point, $\Delta\bm{k}$ as
\begin{eqnarray}
	\epsilon_{\pm}(\zeta,\bm{k}_{\rm BC}+\Delta\bm{k}) \simeq \epsilon_0(\Delta\bm{k}) \pm \sqrt{m^2 + v_{\pm}^2 |\Delta\bm{k}|^{2n}},
	\label{eq:EffBandGen}
\end{eqnarray}
where a ``mass" $m$ is given as $m=c_{\alpha}V_{\alpha}\zeta$ with a coefficient $c_{\alpha}$.
In the following sections, the location of the band crossing point $\bm{k}_{BC}$, dispersion $\epsilon_0 (\Delta\bm{k})$, coefficient $c_{\alpha}$, and ``velocities" $v_{\pm}$ will be determined from the parameters of the microscopic hamiltonians for  several choices of lattice structures.

At $T=0$, the free energy has a simple form,
\begin{eqnarray}
	f[\zeta]-f[0]&\simeq& \frac{2n_{\rm BC}\Omega_{d}}{V_{\rm BZ}}Q + z_{\alpha}n_{\rm u}V_{\alpha}\zeta^2,
\end{eqnarray}
\begin{eqnarray}
	Q=\int_{0}^{\Lambda} k^{d-1}d k \left[ -\sqrt{m^2 + v_{-}^2 k^{2n}} + v_{-} k^n \right]
	\label{Int_Q}
\end{eqnarray}
where $k=|\bm{k}|$, $n_{\rm BC}$ is the number of band-crossing points, $V_{\rm BZ}$ is the volume of the Brillouin zone, and $\Omega_d = 2\pi$ ($4\pi$) for $d=2$ ($d=3$).

The singular part of the free energy arises from
\begin{eqnarray}
	Q = v_{-}|m|^{\frac{d}{n}+1} \int_{0}^{\Lambda/|m|^{\frac{1}{n}}} q^{n+d-1}d q \left[ 1 - \sqrt{1 + \frac{1}{v_{-}^2 q^{2n}}} \right],
	\nonumber \\
	\label{Int_sing}
\end{eqnarray}
where we rescale the integration variable as $q = k/|m|^{1/n}$ to 
eliminate
the singular $m$ dependence
from
the integrand.
Here we note that 
the integral in the right hand side of Eq.(\ref{Int_sing}), 
in the limit $|m| \rightarrow 0+$,
causes a singular part of $Q$.
In fact, by expanding the integrand in power of $1/q^{2n}$
for $|m|/v_{-} \ll 1$, we can obtain a series expansion of $Q$ as
\begin{eqnarray}
Q&\simeq& |m|^{\frac{d}{n}+1}\left[C_0(n,d;v_{-},\Lambda) \right. \notag \\
&+& \left. \sum_{\ell=0}^{\infty}C_{2\ell +1}(n,d;v_{-},\Lambda)
\lim_{\varepsilon \rightarrow 0}
\frac{|m|^{2\ell +1 + \varepsilon -\frac{d}{n}} -1}{d-(2\ell +1 +\varepsilon )n}\right], \notag \\
\end{eqnarray}
where the coefficients $C_{0}$ and $C_{2\ell +1}$ 
depend on $n,d,v_{-}$ and $\Lambda$. 
The expansion of $Q$ naturally yields a singular expansion in terms of $\zeta$
depending on the choice of $(n,d)$:
The expansion of $Q$ has a term proportional to $\ln |\zeta|$ for $d=(2\ell +1)n$ and
$|\zeta|^{\frac{d}{n}+1}$ for other choice of $(n,d)$.
Precise expressions of $Q$ for arbitrary amplictude of $|m|$
are given for physically important sets of $(n,d)$ in the following paragraphs.

For the process of the calculation, it is convenient to rescale the integration variable of Eq.(\ref{Int_Q}) as $x = v_{-}k^n$, from which we obtain
\begin{eqnarray}
	Q = \frac{1}{n}v_{-}^{-\frac{d}{n}}\int_{0}^{v_{-}\Lambda^n}x^{\frac{d}{n}-1}dx \left[-\sqrt{m^2 + x^2} + x \right].
	\label{Int_cal}
\end{eqnarray}

For example, when $(n,d)=(1,2)$, namely, for $d/n+1 = 3$, we obtain
\begin{eqnarray}
	Q &=& \frac{1}{nv_{-}^{2}}\int_{0}^{v_{-}\Lambda^n}xdx \left[-\sqrt{m^2 + x^2} + x \right] \notag \\
	&=& \frac{1}{3nv_{-}^{2}}\left[ v_{-}^3\Lambda^{3n} - \left( v_{-}^2\Lambda^{2n} + m^2 \right)^{\frac{3}{2}} + |m|^3 \right],
\end{eqnarray}
and therefore
\begin{eqnarray}
	f[\zeta]-f[0] \simeq a \zeta^2 + b |\zeta|^3 + ({\rm higher\ order}).
\end{eqnarray}
This expansion has a singularity at $\zeta = 0$, which is quite distinct from conventional LGW type expansion.
(see also Sec. \ref{sec:Honeycomb lattice} for example)

For $d=(2\ell +1)n$, the integral Eq. (\ref{Int_cal}) yields a logarithmic correction proportional to $\ln |\zeta|$.
Indeed, for $n=d$, we obtain
\begin{eqnarray}
	Q &=& \frac{1}{nv_{-}}\int_{0}^{v_{-}\Lambda^n}dx \left[-\sqrt{m^2 + x^2} + x \right] \notag \\
	&=& \frac{1}{2nv_{-}}\left[ v_{-}^2\Lambda^{2n} - v_{-}\Lambda^{n}\sqrt{v_{-}^{2}\Lambda^{2n}+m^2} \right. \notag \\
	&-& \left. m^2 \log \left( \frac{v_{-}\Lambda^{n} + \sqrt{v_{-}^{2}\Lambda^{2n}+m^2} }{m} \right) \right],
\end{eqnarray}
which results in
\begin{eqnarray}
	f[\zeta]-f[0] \simeq b \zeta^2 \ln |\zeta| + a\zeta^2 + ({\rm higher\ order})
\end{eqnarray}
(see also Sec. \ref{sec:Kagome lattice} for example).

The band crossing satisfying $d=3n$ or $(n,d)=(1,3)$ is not known in the real material to our knowledge. We, however, predict a general form of the integral as
\begin{eqnarray}
	Q &=& \frac{1}{nv_{-}^3}\int_{0}^{v_{-}\Lambda^n} x^2 dx \left[-\sqrt{m^2 + x^2} + x \right] \notag \\
	&=& \frac{1}{8nv_{-}^3}\left[ 2v_{-}^4\Lambda^{4n} \right. \notag \\
	&-& \left. v_{-}\Lambda^{n}\sqrt{v_{-}^{2}\Lambda^{2n}+m^2}(2v_{-}^2\Lambda^{2n} + m^2) \right. \notag \\
	&-& \left. m^4 \log \left( \frac{v_{-}\Lambda^{n} + \sqrt{v_{-}^{2}\Lambda^{2n}+m^2} }{m} \right) \right],	
\end{eqnarray}
and free-energy expansion as
\begin{eqnarray}
	f[\zeta]-f[0] &\simeq& a\zeta^2 + b \zeta^4 \ln \frac{1}{|\zeta|} + ({\rm higher\ order}).
\end{eqnarray}

When the free-energy expansion does not yield a logarithmic correction, we have
\begin{eqnarray}
	f[\zeta]-f[0] &\simeq& a\zeta^2 + b |\zeta|^{\frac{d}{n}+1} + ({\rm higher\ order}),
\end{eqnarray}
whose minimum is given by
\begin{eqnarray}
	|\zeta| = \left[ - \frac{2a}{b\left(\frac{d}{n}+1\right)} \right]^{\frac{n}{d-n}}.
\end{eqnarray}
This gives the critical exponent $\beta$ defined by $\zeta \propto |a|^{\beta}$ at $\lambda = 0$, $a<0$ as
\begin{eqnarray}
	\beta = \frac{n}{d-n}.
	\label{eq:beta}
\end{eqnarray}
The critical exponent $\delta$ defined by $\zeta \propto |\lambda|^{1/\delta}$ at $a=0$ is obtained from the minimum of
\begin{eqnarray}
	f[\zeta]-f[0] &\simeq& -\lambda \zeta + b |\zeta|^{\frac{d}{n}+1},
\end{eqnarray}
that is,
\begin{eqnarray}
	\zeta = \left[ \frac{\lambda}{b\left(\frac{d}{n}+1\right)} \right]^{\frac{n}{d}}
\end{eqnarray}
and we have
\begin{eqnarray}
	\delta = \frac{d}{n}.
	\label{eq:delta}
\end{eqnarray}
The relationship between $\lambda$ and $\zeta$ is given by
\begin{eqnarray}
	-\lambda + 2a\zeta = 0
\end{eqnarray}
and we have
\begin{eqnarray}
	\zeta = \frac{\lambda}{2a},
\end{eqnarray}
which yields critical exponent $\gamma$ defined by $\partial \zeta / \partial \lambda \propto |a|^{-\gamma}$ as
\begin{eqnarray}
	\gamma = 1.
	\label{eq:gamma}
\end{eqnarray}
These are quite different from conventional-LGW mean-field critical exponents given by $\beta = 1/2, \delta = 3, \gamma = 1$. 

At nonzero temperatures, the first term in Eq.(\ref{genF}), $\ln \left[1+e^{-(\epsilon_{+}(\zeta,\bm{k})-\mu)/T}\right]$, does not vanish.
This leads to the cancellation of singularity appears at $T=0$.
Derivatives of Eq.(\ref{genF}) with respect to $\zeta$,
\begin{eqnarray}
	A = \frac{1}{2}\frac{\partial^2 f}{\partial \zeta^2}[\zeta \rightarrow 0] \notag \\
	B = \frac{1}{4}\frac{\partial^4 f}{\partial \zeta^4}[\zeta \rightarrow 0],
\end{eqnarray}
converge, and we have the free-energy expansion of the conventional LGW form
\begin{eqnarray}
	f[\zeta]-f[0] \simeq -\lambda\zeta + A \zeta^2 + B \zeta^{4} + \mathcal{O}(\zeta^6).
	\label{eq:Ftemp}
\end{eqnarray}

\subsection{Phase transition}

\begin{figure*}[ht]
\centering
\includegraphics[width=14cm]{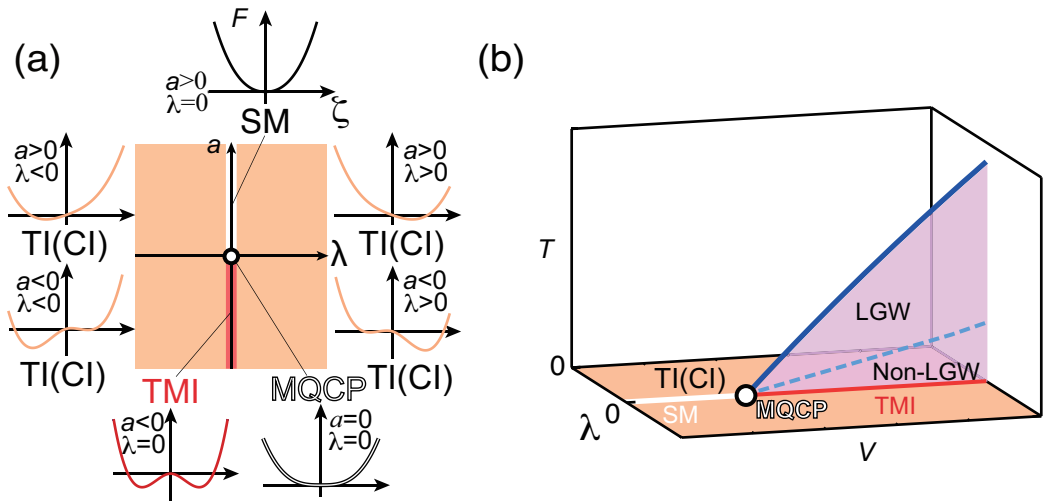}
\caption{
(Color online)
(a) Schematic phase diagram in the plane of $a$ proportional to interaction $V_c-V$ and the field $\lambda$ conjugate to the order parameter $\zeta$. Free energy structures in each phase representing topological insulators (TI), (Chern insulators (CI)), TMI and semimetal (SM) are also shown.
(b) Phase diagram of orbital-current phase induced by Coulomb-repulsion in the space of temperature $T$, $\lambda$ and interaction $V$ obtained by Hartree-Fock calculation.
In the (pink) shaded area at $\lambda=0$, spontaneous orbital current flows and the solid blue line illustrates its critical line.
The dashed blue line shows a crossover between the non-LGW and LGW regions characterized by $\zeta \propto |a|^{n/(d-n)}$, and $\zeta \propto |a|^{1/2}$, respectively \cite{Misawa}.
The white circle at $V=V_{\alpha c}$ and $\lambda=T=0$ is the unconventional quantum critical point (QCP) of the TMT from which the white topological critical line (semimetal) at $T=0$ and $\lambda = 0$ starts unlike the conventional QCP.
\label{fig:Genphase}
}
\end{figure*}

Now in Fig.\ref{fig:Genphase} we illustrate generic features of the phase diagram for OCT obtained from Eq.(\ref{Eq.5}) in the parameter space of the interaction strength $V_{\alpha}$, the symmetry breaking field $\lambda$ and temperature $T$.
Here we summarize basic features of the phase diagram:
A semimetal (SM) is stabilized for a small interaction $V_{\alpha}$ ($\alpha=1,2$) (corresponding to $a>0$ in Eq. (\ref{Eq.5})), at $\lambda=0$ (along the white line in Fig.\ref{fig:Genphase}(b)).
Continuous quantum phase transitions occur immediately into TI or CI when one switches on $\lambda$ either to $\lambda > 0$ or to $\lambda < 0$.
When $V_{\alpha}$ exceeds the critical value $V_{\alpha c}$ (for $a<0$), a spontaneous orbital current (TMI phase) is stabilized for $\lambda = 0$ at $T=0$, where first-order transitions occur when one crosses from $\lambda > 0$ to $\lambda < 0$ through $\lambda = 0$.
The first-order jump terminates at a line of Landau's critical temperature $T_c>0$.
This critical line (blue in Fig.\ref{fig:Genphase}(b)) follows the conventional LGW universality.
This is because the topological nature of the transition is lost at nonzero temperatures and the free-energy expansion recovers the conventional form as shown in Eq.(\ref{eq:Ftemp}).

Sandwiched by Landau's (blue) line at $T_c>0$ and the topological (white) line at $\lambda =0$ and $T=0$, a novel critical point called marginal quantum critical point (MQCP) (white circle) emerges \cite{Imada}.
The critical exponents of the MQCP are generically $\beta = n/(d-n)$, $\gamma=1$ and $\delta=d/n$ as shown in Eqs.(\ref{eq:beta}), (\ref{eq:delta}), and (\ref{eq:gamma}), indicating $\beta\ge 1/2$ and $\delta\le 3$, completely opposite to the standard predictions of LGW type symmetry-breaking transitions, which always satisfy $\beta \le 1/2$ and $\delta\ge 3$.

Though the transition is strictly protected by the topological distinction at $T=0$, large mean-field values for $\beta$ implies an importance of quantum critical fluctuations.
A unified understanding achieved here on the transitions between topologically nontrivial insulators and zero-gap semiconductors tells that the universality is governed by the topological change in the Fermi surface, which is a point in the zero-gap semiconductor and vanishes in the topological insulator.
This is similar to the purely topological ones identified as the marginal quantum criticalities for Mott and Lifshitz transitions.\cite{Imada,Misawa,Imada2,Yamaji,Yamaji2}.
When we compare the three phase transitions; Mott, Lifshitz and the present transitions, the variables for free-energy expansions or the mathematical origins of the singularities are different with each other.
However, in each case, it is common that the dispersion of the electron plays a critical role for the emergence of the unconventionality.

Strong correlation effects emerge as suppressions of simple orders with residual entropy and large quantum fluctuations at low temperatures.  
In the conventional systems with large Fermi surfaces, they emerge typically as momentum and orbital differentiations as found in pseudogap and Fermi arc formation in the cuprate superconductors \cite{Damascelli,Basov}.
In the zero-gap semiconductors, unusual exponents such as large $\beta$ and small $\delta$, resulting in a slow and suppressed growth of the order, indicate the importance of quantum fluctuations, although the transition is strictly protected by the topological distinction at $T=0$.
When we compare the three phase transitions; Mott, Lifshitz and the present transitions, the variables for free-energy expansions or the mathematical origins of the singularities are different with each other.
However, in each case, it is common that the dispersion of the electron plays a critical role for the emergence of the unconventionality.

In this section, we constructed a general theory of the TMT, which predicts a novel singularity of the free energy and resulting critical exponents.
In the following sections, we focus on the specific lattice models and see how we can apply our theory to them.
We numerically confirm the unconventional critical exponents proposed here, and also present deeper understanding about the transitions including those at finite temperatures achieved by theoretical and numerical methods.
Though the lattice models we treat are those with single orbital per site, our theory is applicable to real materials as long as they share the property of the band crossing point.
We will comment on this point in detail in Sec. \ref{sec:Discussion4}.

\section{PHASE TRANSITIONS IN LATTICE MODELS}

\subsection{Honeycomb lattice}\label{sec:Honeycomb lattice}
\begin{figure}
\centering
\includegraphics[width=8.5cm]{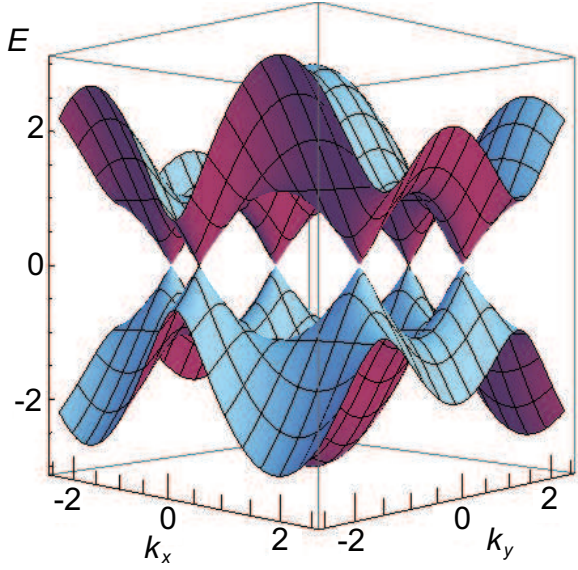}
\caption{
(Color online)
The band structure of $H=H_{0}$ on the honeycomb lattice.
Six Dirac cones shown in the figure are linked to either $K$ or $K'$ by reciprocal lattice vectors.}
\label{honeycombband}
\end{figure}

We first study the phase transition in the honeycomb lattice system at half filling on which the quantum Hall effect \cite{Haldane}, the quantum spin Hall effect \cite{Kane2} and  the topological Mott insulator \cite{Raghu} were discussed before.
We will see that the honeycomb lattice corresponds to $(n,d)=(1,2)$ case.
We choose the unit cell vectors to be
\begin{equation}
	\bm{a}_{1}=(2,0),\ \bm{a}_{2}=(-1,\sqrt{3} ).\label{eq:BravaisTriangle}
\end{equation}
Then the reciprocal lattice vectors are given by
\begin{equation}
	\bm{b}_{1}=\left(1,\frac{1}{\sqrt{3}} \right)\pi,\  \bm{b}_{2}=\left(0,\frac{2}{\sqrt{3}}\right)\pi.
\end{equation}
We also define three-nearest neighbor vectors
\begin{equation}
	\bm{d}_{1} = \left(1,\frac{1}{\sqrt{3}} \right), \bm{d}_{2} = \left(-1,\frac{1}{\sqrt{3}} \right), \bm{d}_{3} = \left(0,-\frac{2}{\sqrt{3}} \right),
\end{equation}
and the next-nearest-neighbor vector
\begin{equation}
	\bm{a}_{3}=(-1,-\sqrt{3} )
\end{equation}
for the purpose of the latter use.
The honeycomb lattice with $H=H_0$ has two equivalent Dirac cones on Brillouin zone \cite{CastroNeto} at
\begin{eqnarray}
	K = \left(\frac{1}{3},\frac{1}{\sqrt{3}}\right)\pi,\ K' =  \left(\frac{-1}{3},\frac{1}{\sqrt{3}}\right)\pi,
\end{eqnarray}
as shown in Fig.\ref{honeycombband}.
The Fermi energy at half filling is on the degeneracy point of these Dirac cones.
Two mean fields we consider here are charge-orbital current
\begin{eqnarray}
	\langle c_{j\beta}^{\dagger}c_{i\alpha}\rangle_{NNN} &=& -i\zeta_c\left( \frac{\bm{b}_{ij}\times\bm{d}_{ij}}{|\bm{b}_{ij}\times\bm{d}_{ij}|} \right)_{z}\sigma_{0\alpha\beta}, \notag \\
	\zeta_c &=& \frac{i}{2}\sum_{\sigma}\langle c_{j\sigma}^{\dagger}c_{i\sigma}\rangle_{NNN}\left( \frac{\bm{b}_{ij}\times\bm{d}_{ij}}{|\bm{b}_{ij}\times\bm{d}_{ij}|} \right)_{z}, \notag \\
	\label{eq:MFHoney1}
\end{eqnarray}
and spin-orbital current
\begin{eqnarray}
	\langle c_{j\beta}^{\dagger}c_{i\alpha}\rangle_{NNN} &=& -i\zeta_s\frac{\bm{b}_{ij}\times\bm{d}_{ij}}{|\bm{b}_{ij}\times\bm{d}_{ij}|}\cdot\bm{\sigma}_{\alpha\beta}, \notag \\
	\zeta_s &=& \frac{i}{2}\sum_{\alpha\beta}\langle c_{j\beta}^{\dagger}c_{i\alpha}\rangle_{NNN}\left( \frac{\bm{b}_{ij}\times\bm{d}_{ij}}{|\bm{b}_{ij}\times\bm{d}_{ij}|} \cdot \bm{\sigma} \right)_{\beta\alpha} \notag \\
	\label{eq:MFHoney2}
\end{eqnarray}
where the notation NNN means that the bond is of the next-nearest-neighbor pair.
We note that the usual Fock term $g$ defined in Eqs. (\ref{SOC}) and (\ref{COC}) disappears in the honeycomb lattice, which makes the following calculations easiyer. 
The former mean field Eq.(\ref{eq:MFHoney1}) breaks the time-reversal symmetry, while the latter Eq.(\ref{eq:MFHoney2}) breaks the SU(2) symmetry.
However, as shown below, the free energy is the same at the mean-field level.
We note that the resulting phase is the same as those studied in Ref.\onlinecite{Haldane} and Ref.\onlinecite{Kane2}.
However, since the spin-orbital-current phase breaks the continuous symmetry, there are Goldstone modes in the ordered phase.
Raghu $et$ $al$ suggested that quantum fluctuations associated with these modes lower the ground-state energy of the spin-orbital-current phase compared to the charge-orbital-current phase \cite{Raghu}.
At finite temperatures, however, the breaking of the continuous symmetry is forbidden because the system is two dimensional, and only the transition to the charge-orbital-current phase is possible.
This fact would modify the phase diagram shown in Fig.\ref{fig:Genphase} because the energy after considering larger fluctuation effects is expected to become lower than a simple ordered state.
Therefore, at $T = 0$, a spin-orbital-current phase or a fluctuating state similar to the spin-liquid state would emerge.
At finite temperatures, we note that this fluctuating state would be stabilized as well instead of the charge-orbital-current phase.
Related to it, the transition to the charge-orbital-current phase driven by the external field may become of the first order at a certain strength of Coulomb interaction.
This may happen if the fluctuation is relatively small and the charge-orbital-current phase continues to exist as a local minimum of the free energy.
Though this is an interesting topic for future, we here focus on the phase transition calculated from the mean-field approximation. 

After the mean-field approximation, we can calculate the band dispersion and the free energy.
Neglecting the effect of $U$ and $V_{1}$, we obtain the energy of the electron with the wave vector $\bm{k}$ by diagonalizing the matrix
\begin{eqnarray}
	H(\bm{k}) =
	\left( \begin{array}{cc}
	2V_{2}\zeta_s s(\bm{k}) \sigma_{z}  & -t\sum_{i=1}^{3}e^{i\bm{k} \cdot \bm{d}_{i}}\sigma_{0} \\
	-t\sum_{i=1}^{3}e^{i\bm{k} \cdot \bm{d}_{i}}\sigma_{0} & -2V_{2}\zeta_s s(\bm{k}) \sigma_{z} \\
	\end{array} \right),
\end{eqnarray}
with
\begin{eqnarray}
	s(\bm{k}) = \sum_{i=1}^{3}\sin(\bm{k} \cdot \bm{a}_{i})
\end{eqnarray}
in the spin-orbital-current phase and
\begin{eqnarray}
	H(\bm{k}) =
	\left( \begin{array}{cc}
	2V_{2}\zeta_c s(\bm{k})   & -t\sum_{i=1}^{3}e^{i\bm{k} \cdot \bm{d}_{i}} \\
	-t\sum_{i=1}^{3}e^{i\bm{k} \cdot \bm{d}_{i}} & -2V_{2}\zeta_c s(\bm{k}) \\
	\end{array} \right) \otimes \sigma_{0}
\end{eqnarray}
in the charge-orbital-current phase.
In this form, it is self evident that the band dispersion and the mean-field free energy of the two states are same for $\zeta_s = \zeta_c$.
Therefore we just write them as $\zeta$ for simplicity in what follows.
The band dispersion is given by
\begin{eqnarray}
	\pm \epsilon(\bm{k}) = \pm \sqrt{t^2\left|\sum_{i=1}^{3}e^{i\bm{k} \cdot \bm{d}_{i}} \right|^2 + 4V_{2}^{2}\zeta^2 \left( \sum_{i=1}^{3}\sin(\bm{k} \cdot \bm{a}_{i}) \right)^2 }. \label{eq:bandHoneycomb} \notag \\
\end{eqnarray}
It is easy to see $\epsilon(\bm{k}) = 0$ at $\bm{k} = K$ or $K'$ when $\zeta=0$.
Then the free energy per unit cell at $T=0$ is given by
\begin{eqnarray}
	f[\zeta] = -\frac{\sqrt{3}}{\pi^2}\int_{BZ}d\bm{k}\epsilon(\bm{k}) + 12V_{2}\zeta^2 \label{eq:honeycombFree0},
\end{eqnarray}
where BZ means the Brillouin zone.
Near $K$ and $K'$ points, we approximate Eq.(\ref{eq:bandHoneycomb}) as
\begin{eqnarray}
	\epsilon(\bm{k}) \simeq \sqrt{3t^2\Delta k^2 + 27V_{2}^{2}\zeta^2 }, \label{eq:effbandHoneycomb}
\end{eqnarray}
where $\Delta k$ is the distance from the $K$ or $K'$ points.
The correspondence with Eq.(\ref{eq:EffBandGen}) is clear with
\begin{eqnarray}
	\bm{k}_{BC} &=& K \ \mathrm{and} \ K', \quad \epsilon_0(\Delta\bm{k}) = 0, \quad v_{\pm} = \sqrt{3}t, \notag \\
	c_2 &=& 3\sqrt{3}, \quad n=1.
\end{eqnarray}
Then we get the free energy as
\begin{eqnarray}
	f[\zeta] &=& -\frac{2\sqrt{3}}{\pi^2}\int_{0}^{\Lambda}2\pi k dk \sqrt{3t^2k^2 + 27V_{2}^{2}\zeta^2 } + 12V_{2}\zeta^2 \notag \\
			     &=& -\frac{4t}{\pi}\left[ \left(\Lambda^2 + \frac{9V_{2}^{2}\zeta^2}{t^2}\right)^\frac{3}{2} - \frac{27V_{2}^3|\zeta|^3}{t^3} \right] + 12V_{2}\zeta^2 \notag \\
			     &\simeq& -\frac{4}{\pi}t\Lambda^3 - \left( \frac{54 \Lambda V_2^2}{\pi t} - 12V_2 \right)\zeta^2 + \frac{108V_2^3}{\pi t^2} |\zeta|^3, \notag \\
			     \label{eq:effFreeHoneycomb0}
\end{eqnarray}
with $\Lambda$ being the cutoff wavelength.
This corresponds to the size of the Dirac cones in the Brillouin zone.
Equation (\ref{eq:effFreeHoneycomb0}) shows $|\zeta|^3$ singularity which comes from $(n,d) = (1,2)$.
The expansion indicates that the phase transition occurs at 
\begin{eqnarray}
	V_{2c}(T=0)=\frac{2\pi t}{9\Lambda}
	\label{V2c0Honeycomb}
\end{eqnarray}
and $\zeta$ is proportional to $V_{2}-V_{2c}$, which can be explicitly described as
\begin{eqnarray}
	\zeta \simeq \frac{27\Lambda^3}{4\pi^2t}(V_{2}-V_{2c}),
	\label{eq:z-VT=0theory}
\end{eqnarray}
for small $V_2 - V_{2c}$. 

We next study the free energy at finite temperatures.
Since the honeycomb lattice has a particle-hole symmetry, $\frac{\partial F}{\partial \mu} = 0$ is automatically satisfied by setting $\mu = 0$, and the free energy using the effective dispersion, Eq.(\ref{eq:effbandHoneycomb}), is given by
\begin{eqnarray}
	f[\zeta] &=& -\frac{2\sqrt{3}T}{\pi^2}\int_{0}^{\Lambda}2\pi k dk \left[ \log\left( 1 + e^{\sqrt{3t^2k^2 + 27V_{2}^{2}\zeta^2}/T} \right) \right. \notag \\
	&+& \left. \log\left( 1 + e^{-\sqrt{3t^2k^2 + 27V_{2}^{2}\zeta^2 }/T} \right) \right] + 12V_{2}\zeta^2.
	\label{eq:effFreeTHoneycomb}
\end{eqnarray}
Though we can not perform the integral in a compact form, we can expand it as
\begin{eqnarray}
	f[\zeta] - f[0] = A\zeta^2 + B\zeta^4
	\label{eq:ExpandFreeTHoneycomb}
\end{eqnarray}
with
\begin{eqnarray}
	A &=& -\frac{36\sqrt{3}V_2^2T}{\pi t^2} \log\left[\cosh\left(\frac{\sqrt{3}t\Lambda}{2T}\right)\right] + 12V_2 ,\notag \\
	B &=& \frac{243V_2^4}{\pi t^2} \left[ \frac{\sqrt{3}}{4T} - \frac{1}{2t\Lambda}\tanh\left(\frac{\sqrt{3} t\Lambda}{2T}\right) \right].
	\label{eq:ExpandFreeTHoneycomb2}
\end{eqnarray}
The coefficient of the third-order differential vanishes at finite temperatures and that of the forth-order differential diverges at $T \rightarrow 0$.
The temperature dependence of the critical value of $V_2$ determined by $A=0$ is given by
\begin{eqnarray}
	V_{2c}(T) = \frac{\sqrt{3}\pi t^2}{9T\log\left[ \cosh \left( \frac{\sqrt{3} t\Lambda}{2T} \right) \right] }.
	\label{V2cTHoneycomb}
\end{eqnarray}
We note that 
\begin{eqnarray}
	V_{2c}(T \rightarrow 0) = \frac{2\pi t}{9\Lambda},
\end{eqnarray}
which shows the consistency with Eq.(\ref{V2c0Honeycomb}).

\begin{figure}
\centering
\includegraphics{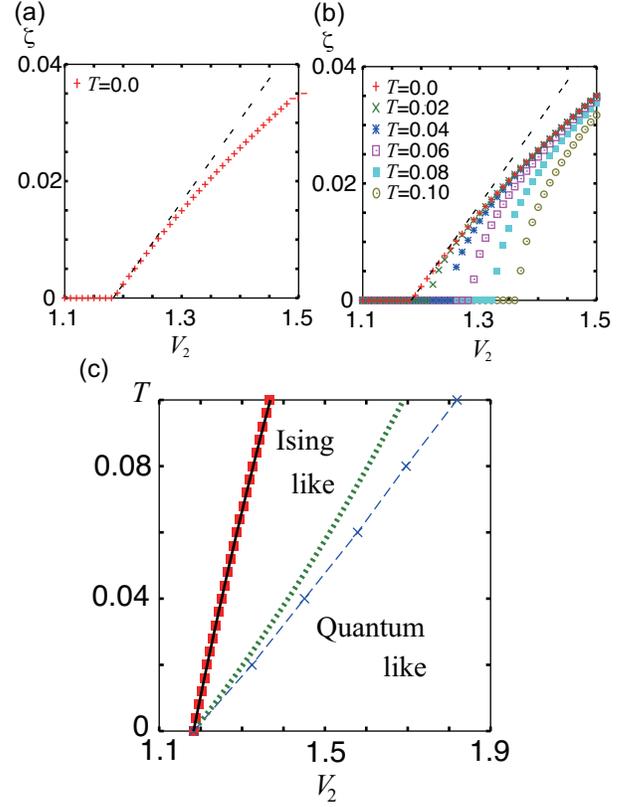}
\caption{
(Color online)
(a) $V_2$ dependence of the order parameter $\zeta$ at $T=0$.
$\zeta$ grows linearly at $T=0$ for small $V_2 - V_{2c}$.
The dashed line shows Eq. (\ref{eq:z-VT=0theory}).
(b) $V_2$ dependence of the order parameter $\zeta$ for several temperatures. $\zeta$ grows with square root of $V_2 - V_{2c}(T)$ at finite temperatures.
(c)The square (red) points show $V_{2c}(T)$ vs. $T$.
They constitute a line which separates the ordered and normal states.
The solid black line is the theoretical critical line defined by Eq.(\ref{V2cTHoneycomb}).
The cross points linked by a dashed line indicate the crossover between Ising-like and quantum-like areas.
The dotted (green) line shows a theoretical expression of the crossover defined by Eq. (\ref{eq:HoneyCrossoverTheo}).
}
\label{fig:HFig1_0}
\end{figure}

Now we compare them with numerical calculations using the band dispersion given by Eq.(\ref{eq:bandHoneycomb}).
Figure \ref{fig:HFig1_0} (a) shows the $V_2$ dependence of the order parameter $\zeta$ at $T=0$ and (b) shows that from $T=0$ up to $T=0.1$.
As discussed above, $\zeta$ grows linearly with $V_2-V_{2c}$ near the critical point at $T=0$.
We find that the critical value of the electron correlation at $T=0$ is $V_{2c}/t \simeq 1.18$, and inserting it to $V_{2c}/t = \frac{2\pi}{9\Lambda}$ gives $\Lambda \simeq 0.590$.
The dashed line in Fig. \ref{fig:HFig1_0} (a) and (b) illustrates Eq. (\ref{eq:z-VT=0theory}) with $\Lambda \simeq 0.590$.
The slope of the dashed line is slightly larger than that of the numerical data because our theoretical expressions are derived from the perfect Dirac cones and neglect the $\bm{k}$ dependence of the orbital current while the linearity itself is preserved.

\begin{figure}
\centering
\includegraphics[width=8.5cm]{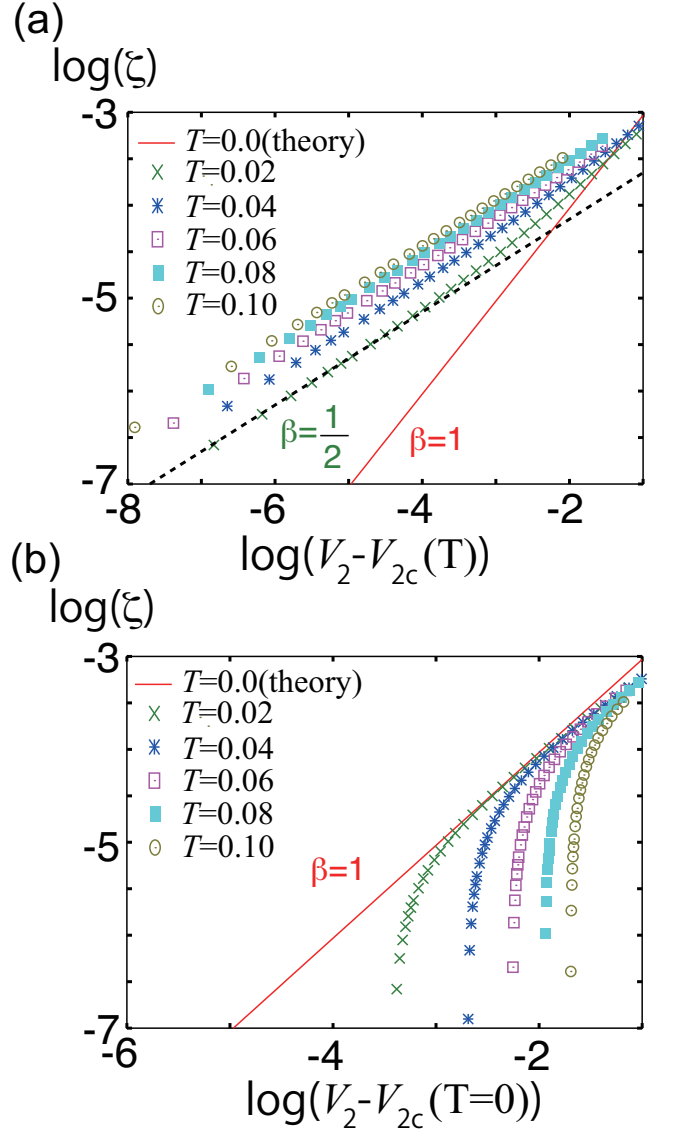}
\caption{
(Color online)
(a) The logarithm of  $\zeta$ and $V_2 - V_{2c}(T)$ for several temperatures.
The slope of the dashed line is given by $\beta=\frac{1}{2}$.
It justifies that $\zeta$ grows with square root of $V_2 - V_{2c}(T)$ for $T>0$.
The solid (red) line is Eq. (\ref{eq:lintheoHoney}).
The crossover value of $V_2$ is exemplified for the case of $T=0.02$ by the intersection of the dotted and solid lines.
(b) The logarithm of $\zeta$ and $V_2 - V_{2c}(T=0)$ showing how Ising-like behaviors reach to the quantum like behavior.
}
\label{fig:HFig1_2}
\end{figure}

At finite temperatures, on the other hand, $\zeta$ is not proportional to $V_{2}-V_{2c}(T)$ and $V_{2c}(T)$ increases with temperature.
Figure \ref{fig:HFig1_0} (c) shows the relation between $V_{2c}(T)$ and $T$, where the square points correspond to the numerical calculations and the (black) solid line is analytical one given by Eq.(\ref{V2cTHoneycomb}) with $\Lambda = 0.590$.
Their agreement indicates that the band dispersion of the honeycomb lattice is expressed well by the Dirac cones by using the effective cut off $\Lambda \simeq 0.590$.
By plotting $\log(\zeta)$ vs $\log(V_2 - V_{2c}(T))$, as shown in Fig. \ref{fig:HFig1_2} (a), we find that the slope of these lines are $\beta = 0.5$ for small $V_2 - V_{2c}(T)$ at $T>0$ which means $\zeta$ is proportional to $\sqrt{V_2 - V_{2c}(T)}$.
This result is consistent with Eq.(\ref{eq:ExpandFreeTHoneycomb}).
As can be seen from Figs. \ref{fig:HFig1_0} (b) and \ref{fig:HFig1_2} (b), $\zeta$ at $T>0$ approaches the line to $V_2(T=0)-V_{2c}(T=0)$ even at finite temperatures for large $V_2$.
We call this region a quantum region.
On the other hand, we call it Ising region when $\zeta$ is proportional to $\sqrt{V_2-V_{2c}(T)}$.
The crossover value of $V_2$ separating the quantum and Ising regions in Fig. \ref{fig:HFig1_0} (c) is calculated as follows:
At $T=0$, numerical fitting shows that the logarithm of $\zeta$ near the critical point is given by
\begin{eqnarray}
	\log(\zeta) = \log\left(V_2 - V_{2c}(T=0)\right) + \log(0.131), \label{eq:lintheoHoney}
\end{eqnarray}
and this is the solid line shown in Figs. \ref{fig:HFig1_2} (a) and (b).
The value $0.131$ comes from the factor of proportionality at $T=0$, which slightly differs from the analytical value $\frac{27\Lambda^3}{4\pi^2} \simeq 0.141$ derived from Eq. (\ref{eq:effFreeHoneycomb0}).
At finite temperatures, $\log(\zeta)$ near $V_{2c}(T)$ is given by
\begin{eqnarray}
	\log(\zeta) = \frac{1}{2}\log \left(V_2 - V_{2c}(T)\right) + \log C(T), \label{eq:lintheoHoneyT}
\end{eqnarray}
where $C(T)$ is a function of the temperature independent of $V_2$.
The dashed line in Fig. \ref{fig:HFig1_2} (a) is this line for the example of $T=0.02$.
Then we define the crossover value of $V_2$ as the intersection of these two lines, which yields
\begin{eqnarray}
	V_2 = V_{2c}(T) + \left( \frac{C(T)}{0.131} \right)^2.
	\label{eq:HoneycombCrossoverCal}
\end{eqnarray}
Data (cross points) corresponding to this equation are shown in Fig. \ref{fig:HFig1_0} (c) with a dashed line, which separates the Ising-like area and quantum-like areas.
We also derive the crossover line from the analytical calculation using Eqs. (\ref{eq:z-VT=0theory}), (\ref{eq:ExpandFreeTHoneycomb}) and (\ref{eq:ExpandFreeTHoneycomb2}), which yields
\begin{eqnarray}
	V_2 = V_{2c}(T) + \frac{2^{6}\pi^{5}t^4}{3^{10}\Lambda^6 V_{2c}^{4} \left[\frac{\sqrt{3}}{2T} - \tanh\left(\frac{\sqrt{3}t\Lambda}{2T}\right)\frac{1}{t\Lambda}\right]}.
	\label{eq:HoneyCrossoverTheo}
\end{eqnarray}
This line with $\Lambda = 0.590$ is shown in Fig. \ref{fig:HFig1_0} (c) as a dotted line.
It is close to the curve Eq.(\ref{eq:HoneycombCrossoverCal}).
A slight difference from the one numerically calculated from the honeycomb lattice is ascribed to the difference between the band structure of the honeycomb lattice and perfect Dirac cones.

Since the linearity of $T$ vs. $V_2$ is good as shown in Fig.\ref{fig:HFig1_0} (c), Eq.(\ref{V2cTHoneycomb}) approximated up to the linear order of $V_2 - V_{2c}$;
\begin{eqnarray}
	T_c \simeq \frac{9\sqrt{3}\Lambda^2}{4\pi\log 2}(V_2 - V_{2c}(T=0))
\end{eqnarray}
holds well with Eq.(\ref{V2cTHoneycomb}).
As discussed above, $\zeta$ is also proportional to $V_2 - \frac{2\pi t}{9\Lambda}$ at $T=0$, and therefore the gap that opens at the Dirac cones has the form
\begin{eqnarray}
	\Delta(T=0) \simeq 3\sqrt{3}V_{2c}\zeta \simeq \frac{9\sqrt{3}\Lambda^2}{2\pi}\left(V_2 - V_{2c}(T=0)\right). \notag \\
\end{eqnarray}
From these expressions, we find that the ratio of $T_c$ and $\Delta(T=0)$ is independent of $V_2$, and
\begin{eqnarray}
	\frac{2\Delta(T=0)}{T_c} \simeq 2\log2 \simeq 1.39.
\end{eqnarray}
The universality of this ratio is similar to the case of BCS superconductivity, $(2\Delta(T=0)/T_c \simeq 3.5)$, while the value is much smaller than the BCS value.
This is because the growth of the gap is suppressed by large $\beta$ compared to the conventional case.

\begin{figure}
\centering
\includegraphics[width=8.5cm]{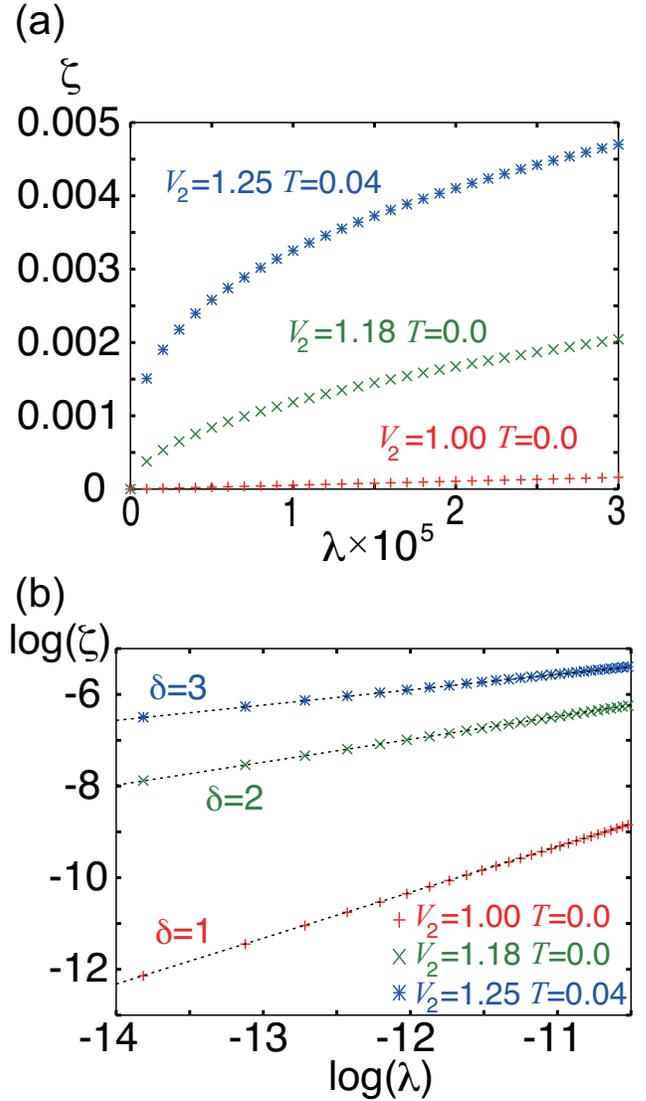}
\caption{
(Color online)
(a) $\lambda$ dependence of $\zeta$ for several values of $V_2$ and $T$.
We note that $V_{2c}(T=0) \simeq 1.18t$ and $V_{2c}(T=0.04) \simeq 1.25t$ for the honeycomb lattice.
(b) Logarithmic plot of data points shown in (a).
The slope of the data in (b) are scaled by $1/\delta$ and can be used to determine $\delta$.
The dotted lines are the slopes expected from the values of $\delta$ written nearby the lines.
This shows the presence of three different criticality;
$\delta = 1$ for $V_2 < V_{2c}(T=0)$ and $T=0$, $\delta = 2$ for $V_2 = V_{2c}(T=0)$ and $T=0$ and $\delta = 3$ for $T>0$.
}
\label{fig:HFig2}
\end{figure}

\begin{figure}
\centering
\includegraphics[width=8.5cm]{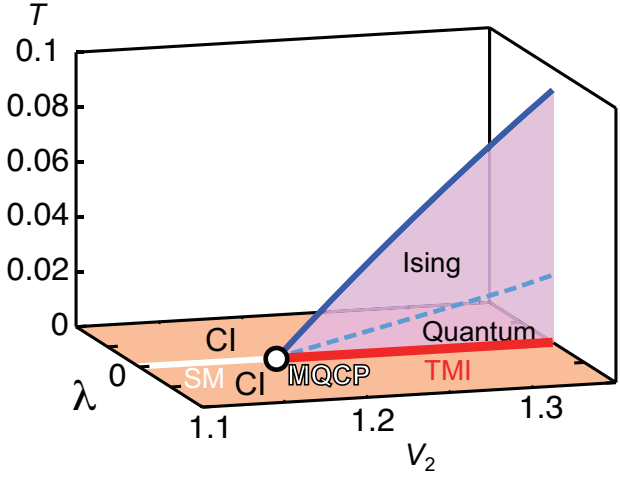}
\caption{
(Color online)
Three dimensional phase diagram of the honeycomb lattice.
In the (pink) shaded surface at $\lambda = 0$, the time-reversal symmetry is spontaneously broken and the solid (blue) line at $T>0$ shows its critical line.
This critical line follows the Ising universality.
The dashed blue line shows a crossover between the quantum and Ising regions characterized by $\zeta \propto |V_2-V_{2c}|$, and $\zeta \propto |V_2-V_{2c}|^{1/2}$.
The white circle at $V_2 \simeq 1.18t$ and $\lambda = T = 0$ is the MQCP.}
	\label{fig:H3Dphase}
\end{figure}

Now we consider the behavior of $\zeta$ under the external field which is conjugate to $\zeta$.
In this case, the conjugate field has the form
\begin{eqnarray}
	i\lambda_c \sum_{\langle \langle ij \rangle \rangle \sigma}\left( \frac{\bm{b}_{ij}\times\bm{d}_{ij}}{|\bm{b}_{ij}\times\bm{d}_{ij}|} \right)_{z}c_{i\sigma}^{\dagger}c_{j\sigma},
\end{eqnarray}
or
\begin{eqnarray}
	i\lambda_s \sum_{\langle \langle ij \rangle \rangle \alpha\beta}\left( \frac{\bm{b}_{ij}\times\bm{d}_{ij}}{|\bm{b}_{ij}\times\bm{d}_{ij}|} \cdot \bm{\sigma} \right)_{\alpha\beta}c_{i\alpha}^{\dagger}c_{j\beta},
\end{eqnarray}
corresponding to Eq. (\ref{eq:MFHoney1}) and Eq.(\ref{eq:MFHoney2}) with $\lambda_c$ and $\lambda_s$ being the control parameter.
They correspond to the periodic magnetic field or the spin-orbit interaction.
Since $\lambda_c$ dependence of $\zeta_c$ and $\lambda_s$ dependence of $\zeta_s$ is completely the same at the mean-field level, we simply abbreviate as $\lambda$.
These add the $\lambda \zeta$ term to the expansion of the free energy.
Therefore, $\zeta$ behave as
\begin{eqnarray}
	\zeta &\propto& \lambda \quad (V_{2}<V_{2c})\notag \\ 
	\zeta &\propto& \lambda^{\frac{1}{2}} \quad (V_{2} = V_{2c}(T=0), \ T=0) \notag \\
	\zeta &\propto& \lambda^{\frac{1}{3}} \quad (V_{2} = V_{2c}(T>0), \ T>0).
	\label{eq:deltaHoneycomb}
\end{eqnarray}
Figure \ref{fig:HFig2} (a) shows the behavior of $\zeta$ with respect to $\lambda$ for each case of Eq.(\ref{eq:deltaHoneycomb}), and (b) is their logarithm.
The overall phase diagram is shown in Fig. \ref{fig:H3Dphase}.
The system becomes topologically nontrivial state at $V_2 > V_{2c}$ and $\lambda =0$ or at $\lambda \neq 0$.
The (white) bold line at $\lambda = 0$ and $V_2 < V_{2c}(T=0)$ in the figure show the critical line of the transition to the semimetal to the topological insulator at $T=0$, and the blue line is the critical line of the spontaneous symmetry breaking.
These lines are connected at MQCP indicated by a white circle.
As shown in Eq. (\ref{eq:deltaHoneycomb}), they have different criticality.

To summarize, our calculations found that the honeycomb lattice indeed shows the novel criticality of $(n,d)=(1,2)$ case proposed in the general theory.
We also gave a theoretical expression as well as numerical results of Landau's critical line at finite temperatures for the honeycomb lattice.

The transition on a diamond lattice follows the same criticality with the honeycomb lattice.
Though the spatial dimensionality is different, the line degeneracy for the diamond lattice around the Dirac dispersion makes the reduction of the effective spatial dimension from 3 to 2.
Therefore the criticality is the same as the honeycomb lattice.
We note that contrary to the honeycomb lattice, the Chern insulator does not exist and only the topological insulator emerges.

\subsection{Kagom\'{e} lattice}\label{sec:Kagome lattice}
\begin{figure}
\centering
\includegraphics[width=8.5cm]{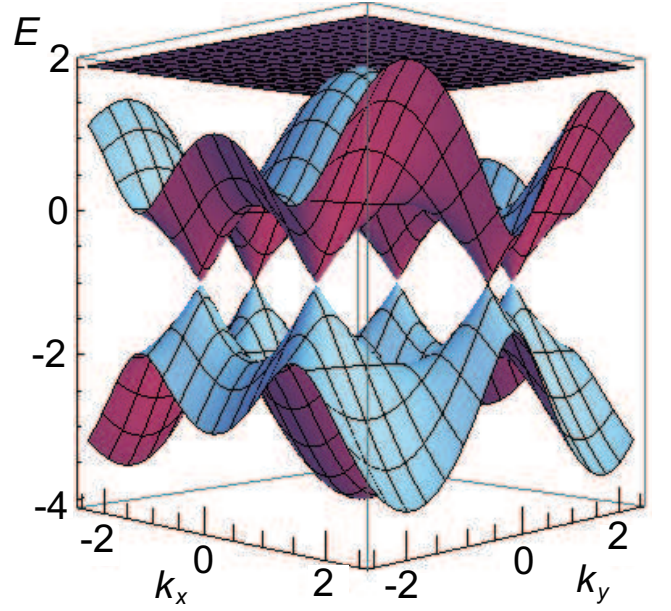}
\caption{
(Color online)
The band structure of the kagom\'{e} lattice.
The highest occupied band at $2/3$ filling touches quadratically to the lowest unoccupied flat band at $\Gamma$ point. 
}
\label{kagomeband}
\end{figure}

\begin{figure}
\centering
\includegraphics[width=8.5cm]{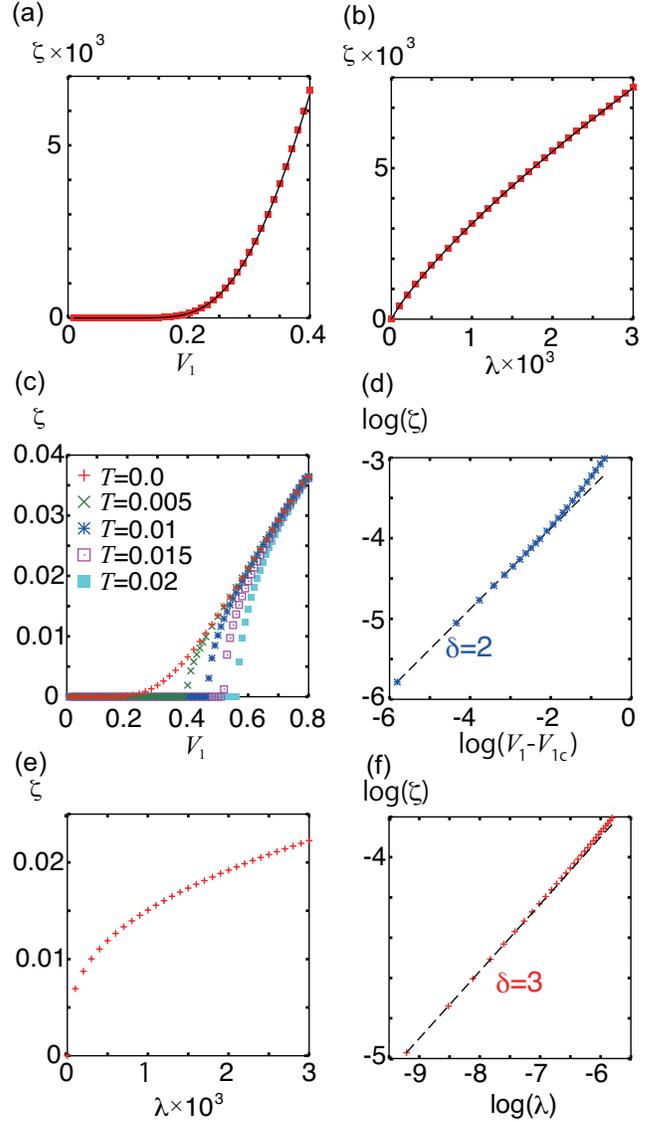}
\caption{
(Color online)
(a) $V_1$ dependence of $\zeta$ at $T=0$ with theoretical curve described by Eq. (\ref{eq:KagomeV-z0}).
(b) $\lambda$ dependence of $\zeta$ at $T=0$ and $V_1 = 0$ with theoretical curve described by Eq. (\ref{eq:Kagomelambda-z0}). $\Lambda = 1.03$ is used in the both cases. The both figures show good agreement of numerical calculations and analytical expressions.
(c) $V_1$ dependence of $\zeta$ for several different choice of temperatures on the kagom\'{e} lattice.
(d) Logarithm of data points at $T=0.01$ and the line with slope $1/\delta=1/2$, which indicate that the criticality belongs to the Ising universality at finite temperatures.
(e) $\lambda$ dependence of $\zeta$ for $T=0.01$ and $V_1=V_{1c}(T=0.01) \simeq 0.467t$.
(f) Logarithm of data points in (a) plotted together with dashed line whose slope is $1/\delta=1/3$ (Ising universality).}
\label{fig:KFig}
\end{figure}

Next we consider the case of the kagom\'{e} lattice at $\frac{2}{3}$ filling.
The TI on the kagom\'{e} lattice has been discussed in several literatures \cite{Liu,Guo2}.
We choose the unit cell vectors to be the same as that of the honeycomb lattice Eq.(\ref{eq:BravaisTriangle}).
In the kagom\'{e} lattice, the mean field of the nearest neighbor bond breaks the time-reversal symmetry or the SU(2) symmetry, which is different from the honeycomb lattice.
We write the mean field of the charge-orbital current and spin-orbital current with $g$ and $\zeta$ as
\begin{eqnarray}
	\langle c_{j\beta}^{\dagger}c_{i\alpha}\rangle_{NN} &=& \left(g-i\zeta_c\left(\frac{\bm{b}_{ij}\times\bm{d}_{ij}}{|\bm{b}_{ij}\times\bm{d}_{ij}|}\right)_z\right)\sigma_{0\alpha\beta}, \notag \\
	\zeta_c &=& \frac{i}{2}\sum_{\sigma}\langle c_{j\sigma}^{\dagger}c_{i\sigma}\rangle_{NN}\left( \frac{\bm{b}_{ij}\times\bm{d}_{ij}}{|\bm{b}_{ij}\times\bm{d}_{ij}|} \right)_{z}. \notag \\
\end{eqnarray}
or
\begin{eqnarray}
	\langle c_{j\beta}^{\dagger}c_{i\alpha}\rangle_{NN} &=& \left(g\sigma_{0}-i\zeta_s\frac{\bm{b}_{ij}\times\bm{d}_{ij}}{|\bm{b}_{ij}\times\bm{d}_{ij}|}\cdot \bm{\sigma} \right)_{\alpha\beta}, \notag \\
	\zeta_s &=& \frac{i}{2}\sum_{\sigma}\langle c_{j\beta}^{\dagger}c_{i\alpha}\rangle_{NN}\left( \frac{\bm{b}_{ij}\times\bm{d}_{ij}}{|\bm{b}_{ij}\times\bm{d}_{ij}|} \cdot \bm{\sigma} \right)_{\beta\alpha}, \notag \\
\end{eqnarray}
where the notation NN means that the bond is of the nearest neighbor pair.
Therefore, the Hamiltonian we consider is Eq.(\ref{Eq.2}) neglecting effect of $U$ and $V_2$.
Defining three nearest neighbor vectors as
\begin{eqnarray}
	\bm{d}_{1} = (1,0), \quad \bm{d}_{2} = \left( \frac{1}{2}, \frac{\sqrt{3}}{2} \right), \quad \bm{d}_3 = \left( -\frac{1}{2}, \frac{\sqrt{3}}{2} \right), \notag \\
\end{eqnarray}
we obtain the mean-field band dispersion by diagonalizing the matrix
\begin{eqnarray}
	H(\bm{k}) &=& -2t'
	\left( \begin{array}{ccc}
	& c_3& c_1 \\
	c_3 & & c_2\\
	c_1& c_2 & \\
	\end{array} \right) \otimes \sigma_0 \notag \\
	&+& 2iV_{1}\zeta_c
	\left( \begin{array}{ccc}
	& -c_3 & c_1 \\
	c_2& & -c_2\\
	-c_1& c_2 & \\
	\end{array} \right) \otimes \sigma_0, \notag \\
	\label{kagomeHk}
\end{eqnarray}
with
\begin{eqnarray}
	c_i = \cos(\bm{k} \cdot \bm{d}_i)
\end{eqnarray}
in the charge-orbital-current phase.
Here, we defined $t' = t + V_{1}g$.
In the spin-orbital-current phase, $\sigma_0$ of the second term is replaced with $\sigma_z$, and again the resulting free energy is the same at the mean-field level.
Therefore we inclusively call them $\zeta$ hereafter.
We note that $g=1/6$ if $\zeta = 0$ and $T=0$.
When $\zeta = 0$, the band has a degeneracy at the $\Gamma$ point $(\bm{k} = (0,0))$ as shown in Fig.\ref{kagomeband}, and the bands touch quadratically there, which makes the transition distinct from the honeycomb lattice.
The dispersions of the highest occupied band and the lowest unoccupied band near the $\Gamma$ point are given by
\begin{eqnarray}
	\epsilon(\Delta\bm{k}) \simeq 2t' -\frac{1}{2}t'\Delta k^2 \pm \sqrt{\frac{1}{4}t'^2\Delta k^4 + 12V_{1}^2\zeta^2}.
\end{eqnarray}
The correspondence with Eq.(\ref{eq:EffBandGen}) is obtained by
\begin{eqnarray}
	\bm{k}_{BC} &=& \Gamma, \quad \epsilon_0(\Delta \bm{k}) = 2t'-\frac{1}{2}t'\Delta k^2, \quad v_{\pm} = \frac{1}{2}t', \notag \\
	c_1 &=& 2\sqrt{3}, \quad n = 2.
\end{eqnarray}
Therefore, the $\zeta$ dependence of the total energy at $T=0$ is given by
\begin{eqnarray}
	&&f[\zeta] - f[0] \notag \\
	&=& -\frac{\sqrt{3}}{\pi^2}\int_{0}^{\Lambda} 2\pi k dk \left( \sqrt{\frac{1}{4}t'^2k^4 + 12V_{1}^2\zeta^2} - \frac{1}{2}t'k^2 \right) \notag \\
	&+& 12V_{1}\zeta^2 \notag \\
	&=&  - \frac{12\sqrt{3} V_{1}^2\zeta^2}{\pi t'}\log\left(\frac{t'\Lambda^2}{4\sqrt{3}V_{1}\zeta} + \sqrt{\frac{t'^2\Lambda^4}{48V_{1}^{2}\zeta^2} + 1}\right) \notag \\
	&-&  \frac{\sqrt{3}\Lambda^2}{4\pi}\sqrt{t'^2\Lambda^4 + 48V_{1}^2\zeta^2} + \frac{\sqrt{3}t' \Lambda^4}{4\pi} + 12V_{1}\zeta^2.
	\label{eq:kagomefree0}
\end{eqnarray}
It is clearly seen that the second-order derivative of Eq.(\ref{eq:kagomefree0}) in terms of $\zeta$ diverges at $\zeta \rightarrow 0$, indicating that a phase transition occurs at $V_1 = 0$.
The self-consistent equation $\partial F/ \partial \zeta = 0$ results in
\begin{eqnarray}
	&&\frac{\sqrt{3}V_{1}}{\pi t'}\log\left(\frac{t'\Lambda^2}{2\sqrt{3}V_{1}\zeta} \right) \notag \\
	&+& \frac{\sqrt{3} V_{1}}{\pi t'} \log \left(\frac{1}{2} + \sqrt{\frac{1}{4} + \frac{12V_{1}^{2}\zeta^2}{t'^2\Lambda^4}}\right) = 1.
	\label{eq:selfconsistentkagome}
\end{eqnarray}
To see the $V_1$ dependence of the order parameter $\zeta$ for small $\zeta$, we neglect the second term of Eq.(\ref{eq:selfconsistentkagome}) since it vanishes at $\zeta \rightarrow 0$.
Then we get an essentially singular behavior as
\begin{eqnarray}
	\zeta = \frac{t\Lambda^2}{2\sqrt{3}V_{1}}e^{-\frac{\pi t}{\sqrt{3} V_1}}.
	\label{eq:KagomeV-z0}
\end{eqnarray}
Here we replaced $t' = t + V_{1}g$ with $t$ since $V_1$ is small.
The (red) data points of Fig. \ref{fig:KFig} (a) obtained by numerically calculating the free energy by diagonalizing Eq.(\ref{kagomeHk}) show $V_1$ dependence of $\zeta$ at $T=0$.
The solid curve shows Eq. (\ref{eq:KagomeV-z0}) with $\Lambda = 1.03$ which very well reproduces the data points.

The singularity is also seen in the field dependence of $\zeta$.
Defining $\lambda$ as the strength of the field conjugate to $\zeta$, we get a theoretical expression
\begin{eqnarray}
	\zeta = \frac{\sqrt{3}\lambda}{\pi t}\log\left(\frac{t\Lambda^2}{4\sqrt{3}\lambda}+\sqrt{\frac{t^2\Lambda^4}{48\lambda^2}+1}\right),
	\label{eq:Kagomelambda-z0}
\end{eqnarray}
at $V_1 = 0$.
We confirm its validity in Fig. \ref{fig:KFig} (b), where the data points of the numerical solution from Eq.(\ref{kagomeHk}) and the theoretical curve Eq.(\ref{eq:Kagomelambda-z0}) with $\Lambda = 1.03$ are shown to agree each other.

Phase transitions at $T > 0$ are complicated compared to the case on the honeycomb lattice because the kagom\'{e} lattice does not have a particle hole symmetry.
This makes the determination of the chemical potential cumbersome.
To avoid this problem, we modify the band dispersion to satisfy the effective particle-hole symmetry which can be accomplished by adding an appropriate third nearest neighbor hopping terms to the Hamiltonian.
This is explicitly described by
\begin{eqnarray}
	H' = -t_3\sum_{\langle \langle \langle ij \rangle \rangle \rangle}(c^{\dagger}_{i\sigma}c_{j\sigma}+h.c.),
\end{eqnarray}
where the third nearest neighbor bond is expressed by $\langle \langle \langle ij \rangle \rangle \rangle$.
Then the effective band energy is given by
\begin{eqnarray}
	\epsilon_{\pm}(\bm{k}) &\simeq& 2t' - 6t_3 -\frac{1}{2}t'k^2 + 6t_3k^2 \notag \\
	&\pm& \sqrt{\frac{1}{4}t'^2k^4 + 12V_{1}^2\zeta^2}.
\end{eqnarray}
By setting $t_3 = t/12$ and neglecting the difference between $t'$ and $t$, we get an effective particle-hole-symmetric dispersion
\begin{eqnarray}
	\epsilon_{\pm}(\bm{k}) \simeq \frac{3}{2}t \pm \sqrt{\frac{1}{4}t^2k^4 + 12V_{1}^2\zeta^2}.
\end{eqnarray}
Then the free energy is given by
\begin{eqnarray}
	f[\zeta] &=& -\frac{\sqrt{3}T}{\pi^2}\int_{0}^{\Lambda}2\pi k dk \left[ \log\left( 1 + e^{-(\epsilon_{-}(\bm{k})-\mu)/T}\right) \right. \notag \\
	&+& \left. \log\left( 1 + e^{-(\epsilon_{+}(\bm{k})-\mu)/T}\right) \right] + 12V_{1}\zeta^2,
	\label{FreeKagomeTgen}
\end{eqnarray}
where $\mu=\frac{3}{2}t$.
We assume that qualitative properties of the phase transition do not change by the approximation above.
This will be confirmed by comparing the universality derived from numerical calculations combined with an approximation in the following way:
After the calculation, we get
\begin{eqnarray}
	f[\zeta] - f[0] = A\zeta^2 + B\zeta^4
\end{eqnarray}
with
\begin{eqnarray}
	A &=& -\frac{12\sqrt{3}V_1^2}{\pi t}\int_{0}^{\frac{\Lambda^2 t}{4T}}\frac{\tanh(x)}{x}dx + 12V_1, \notag \\
	B &=& \frac{36\sqrt{3}V_1^{4}}{\pi t^2T^2}\int_{0}^{\frac{\Lambda^2 t}{2T}} \frac{e^{2x}-1-2xe^x}{x^3(1+e^x)^2}dx.
\end{eqnarray}
Then we reach the relation between $V_{1c}(T)$ and $T$ at low temperatures as
\begin{eqnarray}
	V_{1c}(T) = \frac{\pi t}{\sqrt{3}\int_{0}^{\frac{t\Lambda^2}{4T}}\frac{\tanh x}{x}dx}.
	\label{eq:phslineKagome}
\end{eqnarray}

\begin{figure}
\centering
\includegraphics[width=8.5cm]{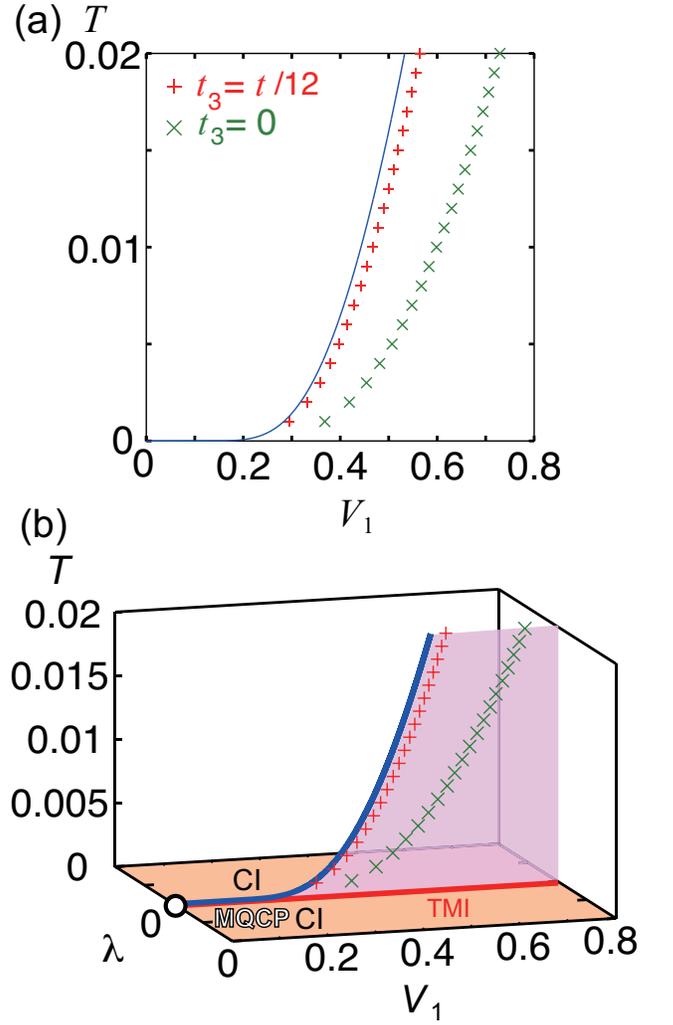}
\caption{
(Color online)
(a) Critical lines of kagom\'{e} lattice with $t_3=t/12$ and $t_3=0$. Solid (blue) line is the analytical expression Eq. (\ref{eq:phslineKagome}) with $\Lambda = 1.03$. The universality is the Ising type at finite temperatures.
Though the numerical results and theoretical expression slightly differ, we can see their qualitative correspondence.
In this case, $T = V_1=0$ is the MQCP, which is a unique feature of $(n,d) = (2,2)$ case.
(b) Phase diagram of kagome\'{e} lattice in the parameter space of $V_1$, $\lambda$, and $T$.
}
\label{fig:KFig4}
\end{figure}

Figure \ref{fig:KFig}(c) shows the $V_1$ dependence of $\zeta$ for several different choice of temperatures, and (b) shows the logarithm at $T=0.01$, which shows that the $\zeta$ grows with square root of $V_1 - V_{1c}(T)$ at finite temperatures.
This is the same as the case of the honeycomb lattice.
External field dependence at $V_1=V_{1c}(T)$ and $T>0$ is also the same as the honeycomb lattice.
That is,
\begin{eqnarray}
	\zeta \propto \lambda^{\frac{1}{3}} \quad (V_{1} = V_{1c}(T), T>0)
\end{eqnarray}
is satisfied as shown in Fig. \ref{fig:KFig}(e) and (f), where the $\lambda$ dependence of $\zeta$ with its logarithm at $T=0.01$ and $V_1 = V_{1c}(T=0.01) \simeq 0.467t$ is shown.
The critical line separating the ordered and normal phases are shown in Fig. \ref{fig:KFig4}.
The numerical solution for the minimum of the free energy Eq.(\ref{FreeKagomeTgen}) for $t_3=t/12$ and $t_3=0$ are shown as data points, and a theoretical expression Eq. (\ref{eq:phslineKagome}) with $\Lambda = 1.03$ for $t_3 = t/12$ is shown for comparison as the solid curve.
The behavior of points and the curve Eq. (\ref{eq:phslineKagome}) have qualitative similarity , while they are still quantatively different compared to the case in the honeycomb lattice Fig.\ref{fig:HFig1_0} (c).
This is because the dispersion of the kagom\'{e} is particle-hole symmetric only at the vicinity of $\Gamma$ point even at $t_3 = t/12$.
However the qualitative equivalence shows the validity of our calculations.
We note that the critical line like Eq. (\ref{eq:phslineKagome}) is also seen in the checker board lattice \cite{Sun}, which is consistent because the effective band dispersion near the Fermi level is quadratic as well.

Let us summarize the phase transition of the kagom\'{e} lattice.
It corresponds to the case of $(n,d) = (2,2)$ in our general theory.
A unique point of the present case is that $V_{1c} = 0$ and topological critical line (white line in Fig.\ref{fig:Genphase} (b)) does not exist.
Consequantly the free energy expansion Eq. (\ref{eq:kagomefree0}) leads to an essential singular behavior Eq.(\ref{eq:KagomeV-z0}) and critical exponents are ill defined at $T=0$.
However it recovers the conventional LGW exponents at finite temperatures.
The exponents at the finite temperatures from numerical calculations agree well with those derived from the band structure retaining the perfect particle-hole symmetry, which justifies our treatments.

\subsection{Pyrochlore lattice}\label{sec:Pyrochlore lattice}
As the last example, we study the pyrochlore lattice.
We choose the unit cell vectors to be
\begin{equation}
	\bm{a}_{1}=(2,0,2),\ \bm{a}_{2}=(0,2,2)\  \textrm{and} \ \bm{a}_{3}=(2,2,0).
	\label{eq:Bravais}
\end{equation}
and the vectors to specify four sites in unit cell as
\begin{eqnarray}
	\bm{d}_{1}&=&(1,0,1),\ \bm{d}_{2}=(0,1,1),\ \bm{d}_{3}=(1,1,0)\ \textrm{and} \notag \\
	\bm{d}_{4}&=&(0,0,0) .\label{eq:sublatticevector}
\end{eqnarray}
We consider the spin-dependent mean field given by
\begin{eqnarray}
	\langle c_{j\beta}^{\dagger}c_{i\alpha}\rangle_{NN} &=& \left( g\sigma_{0} - i\zeta\frac{\bm{b}_{ij}\times\bm{d}_{ij}}{|\bm{b}_{ij}\times\bm{d}_{ij}|}\cdot\bm{\sigma} \right)_{\alpha\beta} \notag \\
	\zeta &=& \frac{i}{2}\sum_{\alpha\beta}\langle c_{j\beta}^{\dagger}c_{i\alpha}\rangle_{NN} \left(\frac{\bm{b}_{ij}\times\bm{d}_{ij}}{|\bm{b}_{ij}\times\bm{d}_{ij}|}\cdot\bm{\sigma} \right)_{\beta\alpha}. \notag \\
	\label{eq:TMI_bond}
\end{eqnarray}
Mean field electron band is given by diagonalizing $H_{0}(\bm{k}) + J_{\zeta}(\bm{k})$, where
\begin{eqnarray}
	H_0(\bm{k}) &=& -2t'
	\left( \begin{array}{cccc}
	0 & c_{xy} & c_{yz} & c'_{zx} \\
	c_{xy} & 0 & c_{xz} & c'_{yz} \\
	c_{yz} & c_{xz} & 0 & c'_{xy} \\
	c'_{zx} & c'_{yz} & c'_{xy} & 0 \\
	\end{array} \right)
	\otimes \sigma_0,
	\label{eq:H0kPyrochlore}
\end{eqnarray}
and
\begin{eqnarray}
	J_{\zeta}(\bm{k}) = \sqrt{2}i\zeta 
	\left(
	\begin{array}{cccc}
	0& u_{xy} &  -u_{yz} & v_{zx}\\
	-u_{xy}&0&u_{zx} & v_{yz}\\
	u_{yz} &-u_{zx} & 0 & v_{xy}\\
	-v_{zx} & -v_{yz} & -v_{xy} & 0\\
	\end{array}
	\right),
\end{eqnarray}
with
\begin{eqnarray}
	t' &=& t+V_1{g} \notag \\
	c_{ij} &=& \cos(k_i - k_j), \notag \\
	c'_{ij} &=& \cos(k_i + k_j), \notag \\
	u_{ij} &=& (\sigma_i + \sigma_j ) \cos (k_i - k_j), \notag \\
	v_{ij} &=& (\sigma_i - \sigma_j ) \cos (k_i + k_j).
\end{eqnarray}

By diagonalizing the mean-field hamiltonian, we obtain 4 bands \cite{Kurita}.
Out of the 4 bands, we focus on 3 bands touching each other at the $\Gamma$ point for $\zeta=0$.
In contrast to simple conduction- and valence-band dispersions already discussed, here, we need to handle 3 bands,
\begin{eqnarray}
	\epsilon = 2t', 2t', 2t'-2t'k^2 + \mathcal{O}(k^4).
\end{eqnarray}
By considering the spin degenracy,
we have 6 bands touching at the band crossing point.

In the presence of non-zero spin-orbital currents $\zeta \neq 0$, 
we obtain 3 split bands,
\begin{eqnarray}
	\epsilon &=& 2t' + 2\sqrt{2}V_{1}\zeta, \notag \\
	&&2t' - \sqrt{2}V_1 \zeta -t' k^2 \pm \sqrt{(\sqrt{2}V_1 \zeta - t' k^2)^2 + 16 V_{1}^2 \zeta^2}. \notag \\
\end{eqnarray}
Each dispersion is doubly degenerate.
At the $\Gamma$ point of the Brillouin zone, $k=0$, in terms of the group theory, 6 degenerate bands in the $T\otimes SU(2)$-manifold split into a doublet ($E_{5/2}$) and quartet ($G_{3/2}$), due to spin-orbital currents.

As it has been already mentioned in Ref.\onlinecite{Guo}, for non-interacting electrons with spin-orbit couplings on pyrochlores, the sign of $\zeta$ is crucial as to whether or not the system becomes topological insulators.
We note that, for the honeycomb and kagom${\rm \acute{e}}$ lattices, topological Mott states appear for both positive and negative $\zeta$. 
However, for the pyrochlore lattice, the sign of $\zeta$ reverses the level of the doublet $E_{5/2}$ and quartet $G_{3/2}$.
Only if the doublet $E_{5/2}$ becomes occupied, the system becomes a topological Mott insulator. 

The free energy at $T=0$ is given by
\begin{eqnarray}
	&&f[\zeta] - f[0] \notag \\
	&\simeq& \frac{32}{\pi^2}\int_{0}^{\Lambda} dk k^2 \notag \\
	&\times& \left(t'k^2 - \sqrt{2}V_{1}\zeta \sqrt{(t'k^2 - \sqrt{2}V_{1}\zeta)^2 + 16V_{1}^2\zeta^2} \right) \notag \\
	 &+& 24V_1\zeta^2.
\end{eqnarray}
From this form of the free energy, we have
\begin{eqnarray}
	\frac{\partial^2 f}{\partial \zeta^2}[\zeta = 0] = -\frac{512 V_{1}^2\Lambda}{\pi^2 t'} + 48V_1
\end{eqnarray}
and
\begin{eqnarray}
	\frac{\partial^3 f}{\partial \zeta^3} &=& \frac{1536t'}{\pi^2\sqrt{|\zeta|}}\left(\frac{V_1}{t'}\right)^{\frac{5}{2}} \notag \\
	&\times&\int_{0}^{\sqrt{\frac{t'}{V_1|\zeta|}}\Lambda}dx \frac{x^6(18-\mathrm{sgn}(\zeta)\sqrt{2}x^2)}{(18-\mathrm{sgn}(\zeta)2\sqrt{2}x^2+x^4)^{\frac{5}{2}}} \notag \\
\end{eqnarray}
from which the energy expansion is given by
\begin{eqnarray}
	f[\zeta] - f[0] = a\zeta^2 + b_{\pm}|\zeta|^{\frac{5}{2}} + \mathcal{O}(\zeta^4),
	\label{eq:PyroFexpansion}
\end{eqnarray}
where
\begin{eqnarray}
	a &=& 24V_1 - \frac{256V_1^2\Lambda}{\pi^2 t'}, \notag \\
	b_+ &=& 9.37t'\left( \frac{V_1}{t'} \right)^{\frac{5}{2}}, \ (\mathrm{for} \ \zeta > 0)\notag \\
	b_- &=& 55.4t'\left( \frac{V_1}{t'} \right)^{\frac{5}{2}}, \ (\mathrm{for} \ \zeta < 0).
\end{eqnarray}
Here, we have numerically calculated
\begin{eqnarray}
	b_{\pm} = \frac{2496t'}{5\pi^2}\left(\frac{V_1}{t'}\right)^{\frac{5}{2}}\int_{0}^{\infty}dx \frac{x^6(18\mp \sqrt{2}x^2)}{(18\mp 2\sqrt{2}x^2+x^4)^{\frac{5}{2}}}. \notag \\
\end{eqnarray}
This form of the free-energy expansion is again non-LGW type where $f[\zeta]$ has a singularity at $\zeta = 0$.

\begin{figure}
\centering
\includegraphics[width=8.5cm]{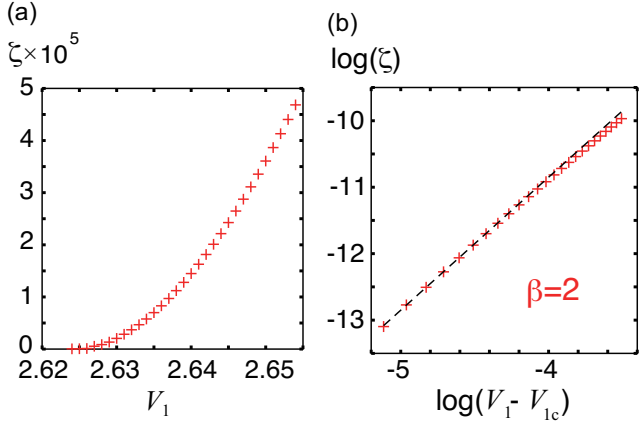}
\caption{
(Color online)
(a) $V_1$ dependence of $\zeta$ at $T=0$ on the pyrochlore lattice.
(b) Logarithmic plot of (a) with the dashed line indicating the slope $\beta = 2$.
An unconventional criticality expected from the singular form of the free-energy expansion Eq.(\ref{eq:PyroFexpansion}) is indeed confirmed here as $\zeta \propto (V_1 - V_{1c})^2$, namely $\beta = 2$.
}
\label{fig:PFig1}
\end{figure}

For small $V_1 - V_{1c}$, we expect that $\zeta$ behaves as $\zeta \propto (V_1 - V_{1c})^2$.
This is confirmed in Fig. \ref{fig:PFig1}, which shows $V_1$ dependence of $\zeta$ at $T=0$.
We find that $V_{1c}\simeq 2.62t$, from which we get $\Lambda = 0.507$.

\begin{figure}
\centering
\includegraphics[width=8.5cm]{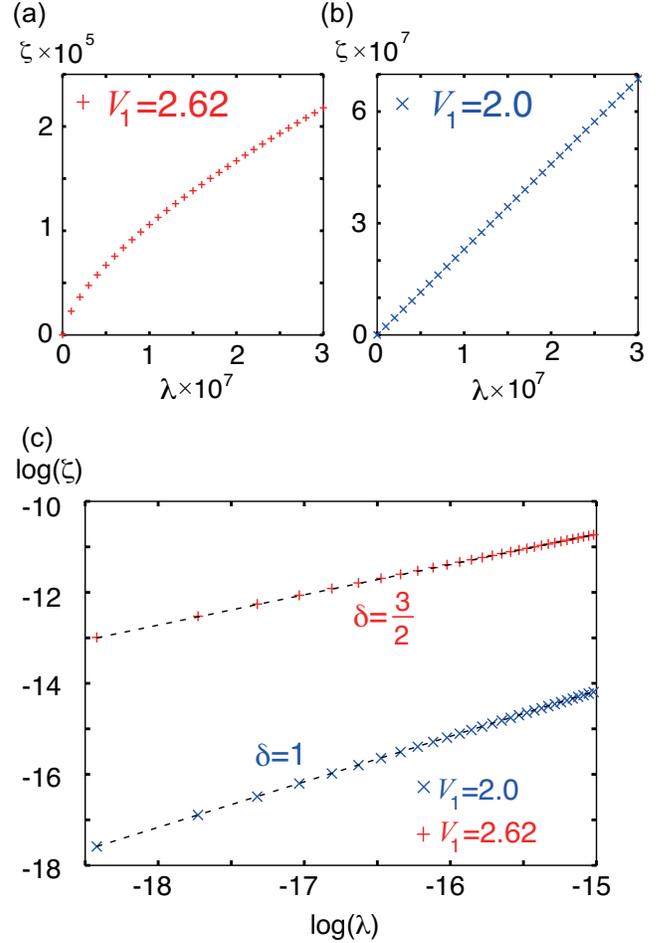}
\caption{
(Color online)
(a) $\lambda$ dependence of $\zeta$ at $V_1=2.62t$ and $T=0$. (b) $\lambda$ dependence of $\zeta$ at $V_1=2.0$ and $T=0$.
(c) Logarithic plot of (a) and (b) with lines of slope given by $1/\delta$.}
\label{fig:PFig2}
\end{figure}

The effect of the $|\zeta|^{\frac{5}{2}}$ term also appears in the $\lambda$ dependence of $\zeta$ at $V_1 = V_{1c}$, where $\lambda$ is the strength of an external field conjugate to $\zeta$.
This corresponds to the spin-orbit interaction described by
\begin{eqnarray}
	 H_{\mathrm{SO}} = \lambda\sum_{\langle ij \rangle \alpha\beta} \left( ic_{i\alpha}^{\dagger}\frac{\bm{b}_{ij}\times\bm{d}_{ij}}{|\bm{b}_{ij}\times\bm{d}_{ij}|}\cdot\bm{\sigma}_{\alpha\beta}c_{j\beta} + h.c. \right), \notag \\
\end{eqnarray}
Since the $\zeta^2$ term disappears at $V_1 = V_{1c}$, $\zeta \propto \lambda^{\frac{2}{3}}$ is satisfied for small $\lambda$.
Figure \ref{fig:PFig2} (a) shows that the $\lambda$ dependence of $\zeta$ at $V_1 = V_{1c}$, and (b) shows the case $V_1 < V_{1c}$.
Their logarithms shown in Fig. \ref{fig:PFig2} (c) fitted with the dotted lines show slopes consistent with the critical exponents $\delta=3/2$ and $\delta=1$ for $V_1 = V_{1c}$ and $V_1 < V_{1c}$, respectively.

At finite temperatures, we find that the transition becomes of first order, which was not observed in previous models.
\begin{figure}
\centering
\includegraphics[width=8.5cm]{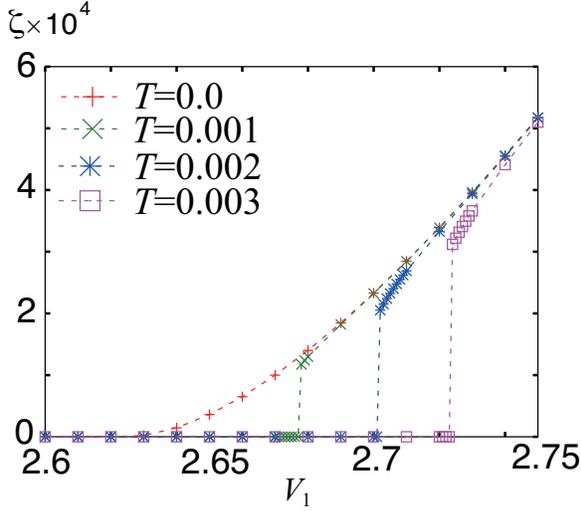}
\caption{
(Color online)
$V_1$ dependence of $\zeta$ for several values of $T$ on the pyrochlore lattice.  $\zeta$ grows with $(V_1 - V_{1c})^2$ at $T=0$. At finite temperature, however, we can see that the transition is of first order, which was not observed in the other lattices.}
\label{fig:PFig3}
\end{figure}
We can see that the jump of the order parameter $\zeta$ increases as a function of $T$.
We note that we added a third neighbor hopping term with $t_3=t/12$ to the Hamiltonian in order to avoid numerical difficulties arising from the two flat bands.
The presence of the $t_3$ term does not change the qualitative properties of the transition.

\begin{figure}[t]
\centering
\includegraphics[width=6cm]{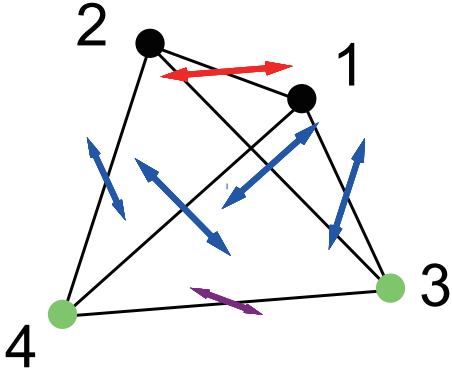}
\caption{
(Color online)
Schematic picture of unit cell of TMI+CDW phase. Black and green sites stand for charge-rich and charge-poor sites, respectively.
Arrows indicating vectors $\bm{v}_{ij}$ in Eq. (\ref{Eq:bondgen}).
}
\label{Fig:TMICDW}
\end{figure}

\begin{figure}[t]
\centering
\includegraphics[width=8.5cm]{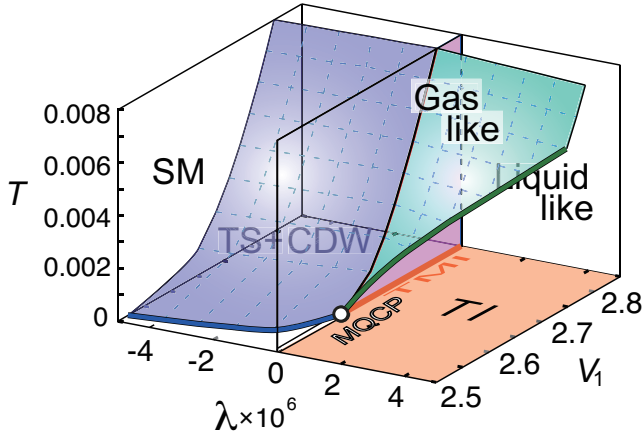}
\caption{
(Color online)
Phase diagram of pyrochlore lattice.
When $\lambda=0$, the system undergoes a transition from SM for small $V_1$ to TMI (dark orange line) at the MQCP, $V_1=V_{1c}\simeq 2.62$t at $T=0$ (white circle).
The MQCP shows a new universality similar to Fig.\ref{fig:Genphase} beyond the conventional LGW scheme. 
When $\lambda < 0$, SMs for small $V_1$ make a first-order transition (the light blue surface), to a topological semiconductor coexisting with charge order (TS+CDW).
The TMI coexiting with charge order (TMI+CDW) at $\lambda\le 0, T=0$ crossovers to TS+CDW at finite temperatures.
It terminates at a QCL (dark blue). For $\lambda > 0$, a topological insulator with a gap is stabilized for all $V_1\ge 0$ at $T=0$.
At $T>0$, through a light green surface, $\zeta$ jumps.
This surface separates a gas-like and liquid-like topological semiconductors (TS) and terminates on the Ising critical (bold green) line, which further terminates at the MQCP.
The MQCP also terminates two QCLs at $T=0$, one along $\lambda=0, V<V_{1c}$ (bold white line) and the other (the dark blue line) representing that between TS+CDW and SM. 
}
\label{fig:Fig3}
\end{figure}

Overall phase diagram of the pyrochlore lattice is exceptional but interesting.
Although the pyrochlore lattice has a unique feature, as discussed above, $f[\zeta]$ follows the same form as Eq. (\ref{Eq.5}) with $f_s(\zeta)=|\zeta|^{5/2}$ and $0<b_+<b_-$.
The obtained phase diagram in Fig.\ref{fig:Fig3} contains several distinct phases;
SM, TMI (at $\lambda = 0, V_{1}>V_{1c}$), TI (at $\lambda>0$) and TMI coexiting with charge order (TMI+CDW) (at $\lambda\le 0, T=0$).
In TMI+CDW phase, we found two charge-rich sites and two charge-poor sites in a unit cell as schematically shown in Fig. \ref{Fig:TMICDW}.
Since the symmetry is lower than the ordinary TMI phase, degrees of freedom of the mean-field increase.
In general, the bond mean field in the unit cell with time-reversal symmetry is given by
\begin{eqnarray}
	\langle c_{j\beta}^{\dagger}c_{i\alpha}\rangle_{NN}  = \left(g_{ij}\sigma_0 - i \bm{v}_{ij}\cdot \bm{\sigma} \right)_{\alpha\beta},
	\label{Eq:bondgen}
\end{eqnarray}
with $g_{ij} = g_{ji}$ and $\bm{v}_{ji} = -\bm{v}_{ij}$.
From the symmetry of the TMI+CDW phase shown in Fig.\ref{Fig:TMICDW}, we have
\begin{eqnarray}
	g_{13} = g_{23} = g_{14} = g_{24}.
\end{eqnarray}
Furthermore, Since 3 symmetric operations defined in Sec. \ref{sec:orbital-current} is intact in bond connecting site 1 and 2, we have
\begin{eqnarray}
	\bm{v}_{12} = \zeta_{12}\frac{\bm{b}_{12}\times\bm{d}_{12}}{|\bm{b}_{12}\times\bm{d}_{12}|}.
\end{eqnarray}
This is the same for $\bm{v}_{34}$.
On the other hand, directions of other vectors can not be determined from the symmetry of the phase.
However, once one vector $\bm{v}_{13}$ is defined, other vectors such as $\bm{v}_{14}$ are determined from the symmetry.
Therefore, we have 9 mean-field parameter for this phase.
They are $g_{12}$, $g_{34}$, $g_{13}$, $\zeta_{12}$, $\zeta_{34}$, 3 components of $\bm{v}_{13}$ and CDW order parameter
\begin{eqnarray}
	\rho = \langle n_{1}\rangle - \langle n_{3} \rangle.
\end{eqnarray}
We note that we added on-site Coulomb interaction with $U=5.5$ in order to avoid trivial CDW phase.
In TMI+CDW phase, the triply degenerated bands at $\lambda = 0$ and $V_{1}<V_{1c}$ completely separate with each other.
However the bottom band is adiabatically connected to the bottom band of TI phase, which indicates the topological nontriviality of this phase.
All the phases meet at the MQCP (white circle), which follows a novel universality similar to the other lattices.
The extension of a critical line at $T = 0$ from the MQCP (white lines in Figs.\ref{fig:Genphase} and \ref{fig:Fig3}) is common protected by the topological nature\cite{Imada} and is fundamentally different from the conventional LGW-type criticality even in itinerant electron systems~\cite{Moriya,Hertz,Millis} that does not have such a critical line and is simply described by $d+z$ dimensional classical LGW transitions, where $z$ is the dynamical exponent.
An important difference of Fig.\ref{fig:Fig3} from the {\it symmetric} case in Fig.\ref{fig:Genphase} is that the SM persists for $\lambda<0$.
In addition, a first-order transition surface (light green surface) (represented by $T=T_{\rm 1st}(\lambda,V_1 )$) bends to $\lambda >0$, which separates the {\it ``topological semiconductor"} phase at $T>0$ and $\lambda>0$ into two: {\it Liquid-like} ($T<T_{\rm 1st}$) and {\it gas-like} ($T>T_{\rm 1st}$) semiconductors.

To summarize the phase transition of the pyrochlore lattice, it has significant difference from our general theory because of the triply degenerated band crossing.
The resultant phase diagram is exceptional for it has several distinct phases as discussed above.
However when we focus on the transition at $T=0$, the criticality is governed by our general theory with $(n,d) = (2,3)$.

\section{Discussion}\label{sec:Discussion4}

We find that our general theory is applicable to many lattice models as discussed in previous sections.
Table \ref{TableI} shows the singularity of each case together with its examples, while Table \ref{TableSI} shows effective band dispersions, parameters, and coefficients for the free energy expansions of lattices discussed in this chapter.

\begin{table*}[th]
\caption{
Singular part, order of transition, and critical exponents in doubly degenerate band-crossing point for several choices of band-dispersion exponent $n$ and the spatial dimension $d$.
First column shows $f_s$, singular parts in $f[\zeta]$.
Second column shows order of transitions determined by following
Ehrenfest's and Lifshitz's classifications.
From 3rd to 5th columns, critical exponents around the marginal quantum critical point (MQCP) of TMT are shown.
For $(n,d)=(1,3)$, exponents $\beta$ and $\delta$ include logarithmic corrections denoted by $\pm 0$, where
$\zeta \simeq \pm \sqrt{|a|/b\ln 1/|a|}$ and
 $\zeta\simeq {\rm sign}(\lambda)\sqrt[3]{|\lambda|}\sqrt[3]{\frac{4b}{3}\ln 1/|\lambda|}$
for $|a|/t \ll 1$ and $|\lambda|/t \ll 1$.
For $(n,d)=(2,2)$, ``essential singular" means 
the instability by
an infinitesimally small $V_{\alpha}/t \ll 1$.
The pyrochlore belongs to (2,3), while it has some uniqueness partly because of the coexisting flat dispersions. 
We note that singularities share similarities to several
antiferromagnetic transitions in zero-gap semiconductors~\cite{Sorella92,Bacsi}.
}
\begin{tabular}{cc|cccccc}
\hline\hline
 &  & $f_{\rm s}[\zeta]$ & order of transition & $\zeta\propto |a|^{\beta}$ & $\zeta \propto$ $|\lambda|^{1/\delta}$ 
& example \\
\hline
Dirac $(n=1)$ & $d=2$ & $|\zeta|^3$ &3&$\beta =1$ & $\delta = 2$ 
%& 1 or 2 
&honeycomb \\
 & $d=3$ & $|\zeta|^4 \ln 1/|\zeta|$ &4& $\beta = 1/2 -0$ %, $\zeta\simeq \pm \sqrt{\frac{|a|}{b\ln 1/|a|}}$ 
& 
$\delta = 3+0$ 
\\
\hline
QBC $(n=2)$ & $d=2$ & $|\zeta|^{2} \ln |\zeta|$ &(essential singular) & $\zeta \simeq \frac{\pm 1}{2vc_{\alpha}V_{\alpha}}
 \exp \left[ - \frac{4vz_{\alpha}n_{\rm u}}{c_{\alpha}^2 V_{\alpha}}\right]$ 
 & 1
& kagom\'{e}\\
& $d=3$ &$ |\zeta|^{5/2}$ &2.5& $\beta=2$ & $\delta = 3/2$ 
& pyrochlore \\
\hline
\hline
\end{tabular}
\label{TableI}
\end{table*}

\begin{table*}[t]
\caption{
Summary of band dispersions, parameters depending on lattice structures, and coefficients
for the free energy expansions. 
}
\begin{tabular}{cc|cc|cc}
\hline\hline
& & honeycomb $(n,d)=(1,2)$ & kagom${\rm \acute{e}}$ $(n,d)=(2,2)$ & pyrochlore $(n,d)=(2,3)$\\
\hline
& filling & one electron per site & 4/3 electrons per site & one electron per site\\
\hline
& $\epsilon (\vec{k})$ & 0 & $2t' -t'k^2 /2$ & -\\
& $c_{\alpha}$ & $3\sqrt{3}$ & $2\sqrt{3}$ & - \\
& $v_{-}$ & $\sqrt{3}t$ & $t'/2$ & - \\
\hline
& $\alpha$ & 2 & 1 & 1\\
& $z_{\alpha}$ & 6 & 4 & 6 \\
& $n_{\rm BC}$ & 2& 1 & 1\\
& $n_{u}$ & 2& 3 & 4\\
& $V_{\rm BZ}$ & $\frac{2\pi^2}{\sqrt{3}}$ & $\frac{2\pi^2}{\sqrt{3}}$ & $\frac{\pi^3}{4}$  \\
\hline
& $a$ & $12V_{2}-\frac{54\Lambda V_{2}^2}{\pi t}$ & $12V_{1}-\frac{6\sqrt{3} V_{1}^2}{\pi t'}\left[1 + 2\ln \left( \frac{t' \Lambda^2}{2\sqrt{3}V_{1}} \right) \right]$ & $24V_1 - \frac{256 V_{1}^2 \Lambda}{\pi^2 t'}$ \\
& $b$ & $ b_{\pm} = \frac{108V_{2}^3}{\pi t^2}$ & $ b_{\pm} = -\frac{12\sqrt{3}V_{1}^2}{\pi t'}$ & $b_{+} = 9.37 t' \left(\frac{V_1}{t'}\right)^{\frac{5}{2}}$  \\
& & & &$ b_{-} = 55.4 t' \left(\frac{V_1}{t'}\right)^{\frac{5}{2}}$ \\
& $\Lambda$ & $0.590$ & $1.03$ & 0.507\\
\hline
\hline
\end{tabular}
\label{TableSI}
\end{table*}

Though the model calculations were performed on the system with a single orbital per site, our theory is applicable to the multi-orbital system as long as they share the properties of the band-crossing point.
For example, the triple-degeneracy in the pyrochlore lattice appears in the band calculation of $\mathrm{Tl_2Ru_2O_7}$ \cite{Ishii}.
Furthermore, we note that this should often occur because degeneracies of the band at high-symmetric points are characterized using group theoretical method.
The triple-degeneracy at the $\Gamma$ point of the pyrochlore lattice discussed in our theory is characterized by $T_{2g}$.
This type of degeneracy often appears in the direct product of basis of $O$-group which characterizes the cubic symmetry as shown in Table \ref{Tablegroup}.
Therefore, when the sites of the unit cell and the structure of multi-orbital in a single site satisfies the conditions shown in the table, $T_{2g}$ degeneracy should be observed.
An example often appears in the condensed matter physic is a triplet of $T_{2g}$ electrons split from $d$-electrons, which are termed as $d_{xy}$, $d_{yz}$ and $d_{zx}$.
When the multi-orbital wavefunctions in a single site are characterized by $T_{2g}$ and the lattice symmetry includes either $A_{1g}$, $E$, $T_{1g}$ or $T_{2g}$, a triple degeneracy of $T_{2g}$ has to be retained because $\{ A_{1g}, E_{g}, T_{1g}, \mathrm{and}\ T_{2g} \} \times T_{2g}$ include $T_{2g}$ as shown in Table \ref{Tablegroup}. Then a zero-gap semiconductor is realized with the Fermi level pinned at the triply degenerate point.
When the SO interaction is further considered, the triple degeneracy splits because the largest representation of the double group in ordinary crystal is of four dimensions.
When the triple degeneracy splits into a doubly degenerate lower-energy state and a nondegenerate upper state, the semimetal may continue to be stable with the Fermi level pinned at the doubly degenerate point and the TI does not emerge if the electron correlation is not considered. 
This is the same as the $\lambda<0$ region in the simple single-orbital model of the pyrochlore lattice, as we considered above.
However, if a nondegenerate lowest-energy state splits off from other states, it may become a topological insulator as in the case of the example we have shown for the $\lambda>0$ region of the single-orbital pyrochlore lattice with only the nearest neighbor transfer.
Even when the double degeneracy remains in the lower-energy level, a further lift of the degeneracy by electron correlations into the nondegenerate lowest-energy state may drive the emergence of the topological insulator as in the case of TI+CDW in the $\lambda<0$ region of the single-band example shown in Fig.\ref{fig:Fig3}.

However, we note that, in the both cases with the nondegenerate lowest-energy state at the $\Gamma$ point, whether the resulting phase is topologically nontrivial or not depends on the property of the wave functions away from the $\Gamma$ point and the way of connecting the dispersions to other TRIM, that requires calculations of the band structure and the parity symmetry for the occupied states at other TRIM.

\begin{table*}[t]
\centering
\caption{
Direct products of representations for the $O$-group and their irreducible representation from page 46 of Ref.\onlinecite{cit:Sugano}.
For the $O_h$-group, suffices $g$ and $u$ should be attached to the irreducible representations according to the rules $g \times g = u \times u = g, \ g \times u = u$.
}
\begin{tabular}{c|ccccc}
\hline \hline
& $A_{1}$ & $A_{2}$ & $E$ & $T_{1}$ & $T_{2}$ \\
\hline
$A_{1}$ & $A_1$ & $A_2$ & $E$ & $T_1$ & $T_2$ \\
$A_{2}$ & & $A_1$ & $E$ & $T_2$ & $T_1$ \\
$E$ & & & $A_1 + A_2 + E$ & $T_1 + T_2$ & $T_1 + T_2$ \\
$T_{1}$ & & & & $A_1 + E + T_1 + T_2$ & $A_2 + E + T_1 + T_2$ \\
$T_{2}$ & & & & & $A_1 + E + T_1 + T_2$\\
\hline \hline
\end{tabular}
\label{Tablegroup}
\end{table*}

Though the quantum critical point is lost if  the first-order transition takes place at $T=0$, and the first-order transition is indeed not predicted in the honeycomb or kagom\'{e} lattices, the pyrochlore lattice shows it at finite temperatures.
It is intriguing to associate the first-order transitions between gas- and liquid-like topological semiconductors (for $\lambda>0$) as well as between SM and TMI+CDW (for $\lambda<0$), to puzzling metal-semiconductor transitions in many pyrochlore compounds with bad metallic behavior as found in Tl$_2$Ru$_2$O$_7$~\cite{Ishii} and $Ln_2$Ir$_2$O$_7$\cite{Matsuhira} for several lanthanoid elements $Ln$. 
Phase diagrams of $Ln_2$Ir$_2$O$_7$ have been examined by several authors \cite{Pesin,Wan,Yang,Krempa,Kargarian}.
Here, we have ignored the orbital degeneracy and orbital dependent anisotropy of transfers of the $d$ bands, while in many cases basic structures at the band crossing are preserved even with this complexity.
Indeed, triply degenerate bands at the $\Gamma$ point near the Fermi level shown in the local density approximation \cite{Ishii} supports the relevance of the present general theory.
Whenever zero-gap semiconductors are found, our general scheme for $(n,d)$ applies.

A nonzero spin-orbital current $|\zeta|\neq 0$ in the TMI breaks the rotational SU(2) invariance of the global spin quantization axis preserved in the original Hamiltonian.
Therefore, the Nambu-Goldstone mode of this order is the spin rotation coupled to the orbital motion. 
We call this new collective excitation, {\it spin-orbiton}. The existence of the spin-orbiton is a way to distinguish TMIs from simple TIs induced by the spin-orbit couplings.
Detecting whether such an excitation exists is an experimental challenge.
In the SM phase, diverging spin-orbiton fluctuations may couple to particle-hole excitations resulting in an overdamped mode, and alter the critical dynamics (dynamical exponent)~\cite{Hertz}. 
This is an issue to be pursued.

Furthermore, spin-orbital currents that preserve the time reversal and lattice symmetries, are {\it invisible} in many standard experiments.
TMI thus is a potential candidate of ``hidden" orders behind mysterious insulating behaviors, as the orbital-current-induced pseudogap scenario discussed in cuprate physics~\cite{Chakravarty,Varma}.

We also note that our results are derived from mean-field approximation and the critical exponents themselves may be subject to change.
However, we here discuss that the universality classes remain unconventional even beyond the mean-field theory, by using the hyperscaling relation and a realistic assumption.
We define a critical exponent $\nu$ from the correlation length $\xi$ as
\begin{eqnarray}
	\xi \propto \left| \frac{T-T_c}{T_c} \right|^{-\nu},
\end{eqnarray}
and a critical exponent $\eta$ from the correlation function $G(r)$ at the critical point and the dynamical exponent $z$ as
\begin{eqnarray}
	G(r) \propto r^{-(d+z-2+\eta)}.
\end{eqnarray}
Beyond the trivial mean-field theory, $1/2 < \nu  $ and $\eta > 0$ are satisfied in most models including Ising, XY and Heisenberg\cite{Pelissetto}.
Therefore, we assume these two relations.
Now we express $\beta$ using the hyperscaling as
\begin{eqnarray}
	\beta = \frac{\nu(d+z-2+\eta)}{2} > \frac{d+z-2}{4}
\end{eqnarray}
To evaluate dynamical exponent $z$, we estimate the polarization function
\begin{eqnarray}
	\Pi^{(R)}(\bm{q},\omega) = \frac{1}{N_u} \sum_{\bm{k}}\frac{f(\epsilon_{+}(\bm{k})) - f(\epsilon_{-}(\bm{k}+\bm{q}))}{\omega + i\eta - \epsilon(\bm{q}) + \epsilon_{-}(\bm{k}+\bm{q})}, \notag \\
\end{eqnarray}
because $z$ is given by the scaling ratio of $q$ and $\omega$.
In two-dimensional case, for example, this comes from a microscopic effective Hamiltonian
\begin{eqnarray}
	H_0 &=& -t\sum_{\bm{k}}
	\bm{c}_{\bm{k}}^{\dagger}
	\left( \begin{array}{cc}
	& f(\bm{k}) \\
	f^{*}(\bm{k})& \\
	\end{array} \right)
	\bm{c}_{\bm{k}}
\end{eqnarray}
with $\bm{c}_{\bm{k}}^{\dagger} = (c_{\bm{k}1}^{\dagger}\ c_{\bm{k}2}^{\dagger})$ and the operator of the order parameter
\begin{eqnarray}
	\zeta_{\bm{q}} &=& \sum_{\bm{k}}
	\bm{c}_{\bm{k}}^{\dagger}
	\sigma_z
	\bm{c}_{\bm{k}}.
\end{eqnarray}
Here,
\begin{eqnarray}
	f(\bm{k}) = k_x - ik_y
\end{eqnarray}
for the Dirac cone and
\begin{eqnarray}
	f(\bm{k}) = -k_x^2 + k_y^2 -2ik_xk_y
\end{eqnarray}
for the quadratic band crossing.
In a pyrochlore lattice, the microscopic effective model is slightly complicated because of the 3-fold degeneracy, and the Hamiltonian is given by
\begin{eqnarray}
	H_0 &=& -t\sum_{\bm{k}}
	\bm{c}_{\bm{k}}^{\dagger}
	%\left( \begin{array}{ccc}
	%c_{\bm{k}1}^{\dagger} & c_{\bm{k}2}^{\dagger} & c_{\bm{k}3}^{\dagger} \\
	%\end{array} \right) 
	\left( \begin{array}{ccc}
	k_x^2 & k_xk_y & k_xk_z \\
	k_xk_y & k_y^2 & k_yk_z \\
	k_xk_z & k_yk_z & k_z^2 \\
	\end{array} \right) \otimes\sigma_0
	%\left( \begin{array}{c}
	%c_{\bm{k}1}\\
	%c_{\bm{k}2}\\
	%c_{\bm{k}3}
	%\end{array} \right),
	\bm{c}_{\bm{k}},
\end{eqnarray}
with $c_{\bm{k}}^{\dagger} = (c_{\bm{k}1\uparrow}^{\dagger} \ c_{\bm{k}1\downarrow}^{\dagger}\ c_{\bm{k}2\uparrow}^{\dagger} \ c_{\bm{k}2\downarrow}^{\dagger} \ c_{\bm{k}3\uparrow}^{\dagger} \ c_{\bm{k}3\downarrow}^{\dagger})$.
The operator of the order parameter is given by
\begin{eqnarray}
	\zeta_{\bm{q}} &=& i\sum_{\bm{k}}
	\bm{c}_{\bm{k}}^{\dagger}
	\left( \begin{array}{ccc}
	& -\sigma_z & \sigma_y \\
	\sigma_z &  & -\sigma_x \\
	-\sigma_y & \sigma_x &  \\
	\end{array} \right)
	\bm{c}_{\bm{k}+\bm{q}}.
\end{eqnarray}
Then the correlation function is given by
\begin{eqnarray}
	D^{0}(\bm{q},\nu_{n}) &\equiv& -\int_{0}^{\beta} \langle T_{\tau}[\zeta_{\bm{q}}(\tau)\zeta_{-\bm{q}}] \rangle e^{i\nu_{n}\tau} d\tau \notag \\
	&=& \sum_{\bm{k}uv} \frac{f_{\bm{k}+\bm{q}u} - f_{\bm{k}v}}{i\nu_{n}-(\epsilon_{\bm{k}+\bm{q}u} - \epsilon_{\bm{k}v})}g(\bm{k},\bm{q})_{uv},% \\
\end{eqnarray}
where $u, v$ are band indices, $\epsilon_{\bm{k}u}$ is the band energy calculated from $H_{0}$, $f_{\bm{k}u}$ is the distribution function and $g(\bm{k},\bm{q})_{uv}$ is the coefficient depends on the form of $H_{0}$ and $\zeta_{\bm{q}}$.
In the honeycomb lattice, we have $(n,d) = (1,2)$ and
\begin{eqnarray}
	\Pi^{(R)}(\bm{q},\omega)
	\simeq \begin{cases}
	D_0 + Aq^2 + C\omega^2
  % \ln \frac{\Lambda}{q} 
  & (\omega < vq) \\
	D_0 + Aq^2 + iC' & (vq < \omega)
	\end{cases},
\end{eqnarray}
with $\Lambda$ being a cut-off wave length and $D_0, A, C$ being constants.
Therefore, we have $z=1$ and $\beta > 1/4$. However, as detailed below, $\beta > 1/2$ is supported by numerical simulations.
For example, beyond the mean-field analyses, a quantum Monte Carlo calculation suggests $\beta \sim 0.77$, $\delta \sim 2.3$ and $\gamma \sim 1$ in the semimetal-insulator phase transition of the honeycomb lattice\cite{Drut}, where the simultaneous Fermi-surface topology change and the spontaneous symmetry breaking take place as well. The unconventional mean-field exponents, $\beta=1, \delta=2, \gamma=1$ are only slightly modified. 
In the kagom\'{e} lattice, $(n,d) = (2,2)$ and
\begin{eqnarray}
	\Pi^{(R)}(\bm{q},\omega) \simeq (D_0 + Aq^2) \left[\frac{1}{2v}\ln\frac{v\Lambda}{\omega} + i\frac{v\pi}{2} \right],
\end{eqnarray}
where we can not define $z$.
However, the transition occurs at $V_1=0$ which justify the treatment of the weak coupling limit.
Correspondingly, $\Pi^{(R)}(\bm{q},\omega)$ diverges at $\omega \rightarrow 0$.
Finally, in the pyrochlore lattice, $(n,d) = (2,3)$ and we have
\begin{eqnarray}
	\Pi^{(R)}(\bm{q},\omega) \simeq D_{0} + Aq^2 - iC|\omega|^{1/2}.
\end{eqnarray}
Therefore, $z=4$, which suggests $\beta > 5/4$. Therefore, irrespective of the lattice structure, the unconventional critical exponents are likely even after the fluctuation effects are taken into account beyond the mean-field estimate.
Though the pyrochlore lattice includes flat bands, they do not give serious effects on the estimation of the polarization function, that is, it does not show any divergence even at finite temperatures. This is because when occupied (unoccupied) band is flat, the transition partner of the unoccupied (occupied) band retains quadratic dispersion and the excitation energy keeps quadratic momentum dependence.
Moreover, all of our results related to the pyrochlore lattice do not change under the existence of the additional band dispersion given by
\begin{eqnarray}
	H'(\bm{k}) = t'
	\left( \begin{array}{ccc}
	k^2& &\\
	& k^2 &\\
	& & k^2 \\
	\end{array} \right).
\end{eqnarray}
Actually this comes from further neighbor hopping elements in real space and gives quadratic dispersions to the flat bands.
This also helps us to avoid a divergence which might appear in the renormalization-group approach due to the flat band.

Studies on the nature of the transitions from metals to correlated insulators are one of the hottest open issues in condensed matter physics.
A difficulty in these studies is that the large Fermi surface expected in the noninteracting system has to be seriously modified before opening of the insulating gap.
A typical example is the experimental observation of the Fermi arc or pocket with a pseudogap formation in the underdoped region of the cuprate superconductors.
Theoretically, this complexity has blocked a whole understanding of the topological metal-insulator transitions and leaves an open question how the large Fermi surface evolves into the gapped insulator in the doped Mott insulators.
However, in the present case of the transition to the topological insulator from the zero-gap semiconductors, the metallic side is always described by the Fermi point that does not require solving a difficult issue how the momentum differentiation evolves on the verge of the transition.
From the present example, one can get insights into the nature of the quantum criticality induced by the metal-to-correlated-insulator transition in a simpler situation.
This is indeed true, because, in the present case, the topology-dominated structure of the phase diagram with the QCL together with unusual critical exponents at the marginal quantum critical point have certain similarities to the universality of the Mott transition discussed in the literature (Ref. 5,7,26-28). 
In fact, some of the Mott transitions exemplified in the kappa-ET type molecular conductors have been proposed to have an unconventional critical exponents\cite{KagawaKanoda}.
The exponents and the universality discussed in those studies may well be given better insights with more broad perspective from the present clarifications.

\begin{figure}
\centering
\includegraphics[width=8.5cm]{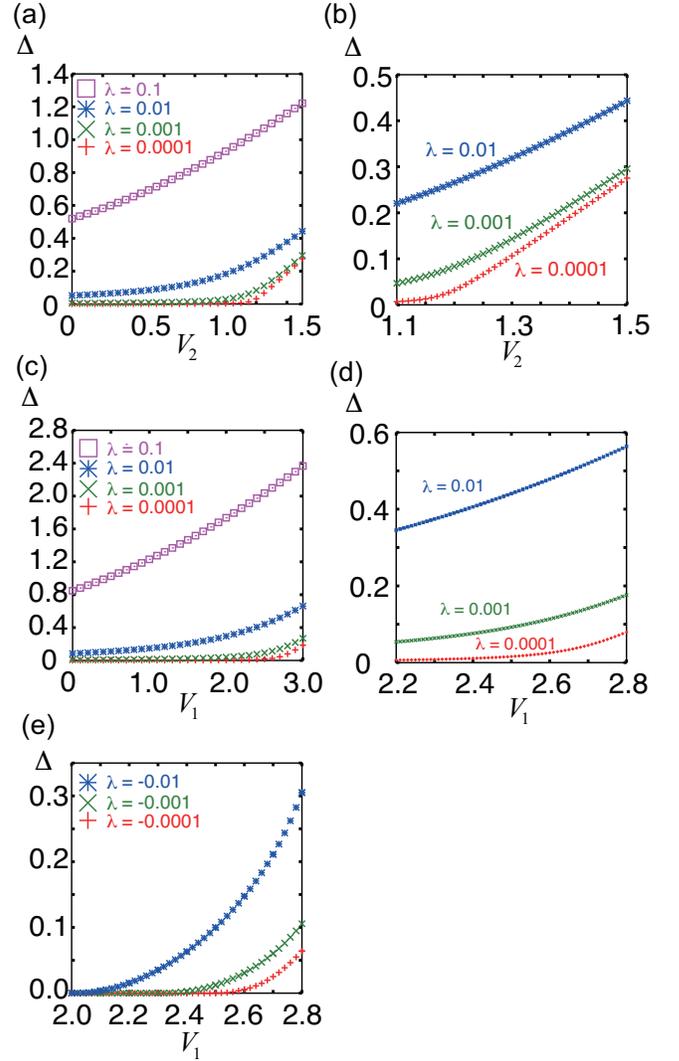}
\caption{
(Color online) 
(a) and (b) Gap enhancement due to the electron interaction in the honeycomb lattice.
(c) and (d) Gap enhancement due to the electron interaction in the pyrochlore lattice.
In each case, significant enhancement of the gap is observed even the interaction does not exceed the critical value for $\lambda = 0$.
(e) Interaction dependence of the gap in the pyrochlore lattice with $\lambda < 0$.
In this case, the external field do not brings the system to the topological insulator, and the phase transition due to the electron interaction is observed.
\label{fig17}}
\end{figure}

In the realistic systems, no matter how small, the spin-orbit interaction $\lambda$ is nonzero.
In particular, $4d$ and $5d$ transition metal compounds have substantial amplitudes. 
In those cases, the intersite interaction we discussed in this paper certainly helps in enhancing the bulk gap amplitude associated with the formation of the topological insulator.
In Fig.\ref{fig17}, we show that the gap is sensitively enhanced even when the interaction does not exceed the critical value for $\lambda=0$. It clearly indicates the role of interaction in the gap of the real materials.

Here we summarize classification of the quantum phase transitions and a category of the quantum phase transitions covered by the present study.
Since the analyses by Landau for symmetry breaking transition and by Lifshitz for Lifshitz transition, Wen readdressed the
category of quantum phase transitions by classifying them into symmetry-breaking and topological
transitions\cite{XGWen}.
The symmetry-breaking transition is totally described by the LGW scheme while
the topological transition is not.
The changes in Fermi-surface topology and emergence of the nontrivial $Z_2$ or $Z$ topological invariants
are typical examples of the topological transitions.

\begin{figure}
\centering
\includegraphics[width=9.5cm]{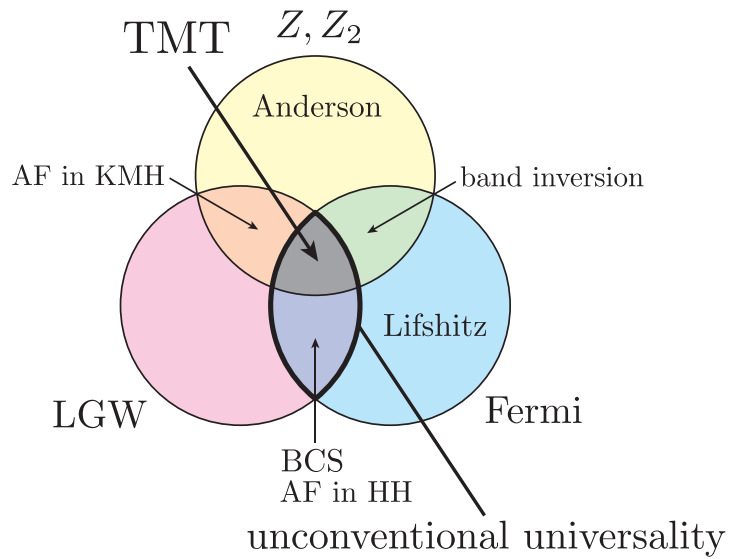}
\caption{
(Color online) Classification of quantum phase transitions.
Three circles, ``LGW", ``Fermi" and ``$Z$,$Z_2$", stand for the symmetry-breaking,
topological changes of the Fermi surface and topological changes of the integer index characterizing the system, respectively.
Lifshitz transitions are
typical examples of quantum phase transitions involving the topological changes in the Fermi surface only, denoted as ``Lifshitz."
Disorders and resultant Anderson localizations offer examples of changes in topological invariants without involving Fermi-surface topology changes
and symmetry breakings, denoted as ``Anderson"\cite{JianLi}.
Overlaps among these three circles offer various qunatum phase transitions.
For example, ``AF in KMH" stands for the antiferromagnetic transitions in the Kane-Mele-Hubbard model\cite{Hohenadler}.
Band inversion transitions between topologically trivial and nontrivial insulators are denoted by ``band inversion."\cite{Murakami1} 
The unconventional universality focused in the present paper appears within the overlap of ``LGW" and ``Fermi."
This class of quantum phase transitions contains the BCS and AF transitions in the honeycomb Hubbard model, denoted as ``AF in HH."
The TMT has all these three characters (``LGW," ``Fermi," and ``$Z$,$Z_2$").
\label{fig:PSBT}}
\end{figure}

The present paper focuses on unconventional quantum criticalities of TMTs
characterized by simultaneous symmetry breakings and topological changes in the Fermi surface together
with emergence of nontrivial $Z$ or $Z_2$ topological invariants
%and an unconventional universality 
(see the classification of quantum phase transitions in Fig.\ref{fig:PSBT}).
Although the present studies on the unconventional criticalities are motivated by growing interest in
correlation-induced topological phases,
it shows that the unconventional universality contains a wider range of quantum phase transitions. 
For example, even when the OCT lacks the emergence of the nontrivial topological invariant,
the OCT necessarily belongs to the unconventional category 
%Especially, the OCT necessarily belongs to the unconventional category
% (SB+T in Fig.\ref{fig:PSBT}).
(overlap of ``LGW" with ``Fermi," representing topological changes in Fermi surface, in Fig.\ref{fig:PSBT}).
We also note that the well-known examples of quantum phase transitions with singular free energy expansions,
for example, antiferromagnetic (AF) transitions in the honeycomb Hubbard model\cite{Sorella92} and the BCS transition,
also belong to the unconventional category.
However, it has not been properly appreciated before that
the topological character with the extension of the
QCL protects the unconventional quantum criticality and is shared by a wide class of quantum phase transitions,
including OCTs, BCS and magnetic transitions in zero-gap semiconductors.

The BCS transition does not have a well established quantum critical line, because its quantum critical line exists in the side of the repulsive effective interaction, which is still highly controversial whether the superconductivity disappears or not.
Furthermore, its criticality
is expressed by the essential singularity,
which can not be described by finite critical exponents and belongs rather to a special case.
As a result, there is no significant impact of the QCL on the quantum criticality of the BCS transitions.

On the contrary, the present study on the OCTs sheds light on the topological protection for the unconventional criticalities.
In addition, the OCTs offer much wider realization of the unconventional criticalities than the BCS transitions and magnetic ones in zero-gap semiconductors.
The OCTs and TMTs accompanied with changes of $Z_2$ of $Z$ topological invariants,
contain a quite general mechanism which results in topology changes of Fermi surfaces by continuous transitions.
This feature is distinct from that in AF transitions, which becomes accidentally unconventional for the honeycomb Hubabrd model.
Therefore, orbital currents in zero-gap semiconductors offer general grounds for studying such unconventional quantum
criticalities by electron correlations.

\section{Conclusion and outlook}
In this paper, we have studied OCT in zero-gap semiconductors.
We show that unconventional phase transitions occur in the topological Mott insulator because of the involvement of the topological change of the Fermi surface, which also leads to a free-energy singularity $|\zeta|^{d/n+1}$.
It generates unconventional universalities characterized by mean-field critical exponents $\beta = n/(d-n)$ and $\delta = d/n$.
We have presented a general theory for doubly-degenerate band crossing and applied to honeycomb, kagom\'{e} lattices.
We have also treated the pyrochlore lattices having triply degenerated band crossing.
Unconventional universalities have been confirmed and in the plane of temperature interaction and spin-orbit interaction (or magnetic fields) phase diagrams for different lattices have been presented using analytical and numerical methods.
We have shown that the OCT inevitably involves the two basic mechanisms of quantum phase transitions, symmetry breaking and topology change, and it yields an unprecedented type
of quantum criticality beyond the conventional LGW scheme characterized by the unusual critical exponents and the emergence of the MQCP as the ending point of the QCL.
Our findings present a solid foundation for unconventional critical phenomena of OCT, and serve a basis for utilization of the OCT such as switching between the topologically distinct phases.
For more quantitative estimate of the critical exponents by considering fluctuations, for example, fermionic renormalization group \cite{Herbut} or a renormalization group method applicable to non-analytic free-energy expansions, which may give a modification to our analysis, will be helpful and are left for future studies.
These may give a solution for dynamical exponents and upper critical dimensions of OCT,
and hence give the estimate for quantitative corrections of the other critical exponents as well.
Nevertheless, the basic structure of the phase diagram has to be protected by the topological structure with the 
emergence of the QCL and the marginal quantum critical point.
Strong quantum fluctuations expected around the present OCT may offer a basis for unprecedented quantum phases mediated by the spin-orbitons.

\section*{Acknowledgement}

The authors thank financial support by Grant-in-Aid for Scientific Research 
(No. 22340090), from MEXT, Japan.
Y.Y. thanks T. Misawa for fruitful discussions.
A part of this
research was supported by the Strategic Programs for Innovative
Research (SPIRE), MEXT, and the Computational Materials Science
Initiative (CMSI), Japan.

\end{document}